\documentclass[reprint,aps,pra,superscriptaddress,twocolumn,showpacs]{revtex4}
\usepackage{epsfig}
\usepackage{graphics}
\usepackage{amsmath}
\usepackage{subfigure}
\usepackage{color}
\usepackage{float}
\usepackage{soul}
\usepackage[usenames, dvipsnames]{xcolor}

\begin{document}
\title{Enhanced Mobility of Discrete Solitons in Anisotropic Two-Dimensional Waveguide Arrays with Modulated Separations}

\author{U. Al Khawaja}
\email{u.alkhawaja@uaeu.ac.ae}
\author{P. S. Vinayagam}
\affiliation{Department of Physics, United Arab Emirates University, P.O. Box
15551, Al-Ain, United Arab Emirates.} 
\author{S. M. Al-Marzoug}
\affiliation{Physics Department, King Fahd University of Petroleum and Minerals, Dhahran 31261, Saudi Arabia} 

\begin{abstract}
We consider two-dimensional waveguide arrays with anisotropic
coupling coefficients. We show using numerical and variational
calculations that four stationary soliton types exist:
Site-Centered, Bond-Centered, Hybrid-$X$ and Hybrid-$Y$. For the
isotropic case the last two modes become identical and equivalent
to the known hybrid soliton. With a variational calculation using
a gaussian trial function and six variational parameters
corresponding to the soliton's position, width, and velocity
components, the four stationary soliton types are reproduced and
their equilibrium widths are accounted for accurately for a wide
range of anisotropy ratios. We obtained using the variational
calculation the Peierls-Nabarro potential and barrier heights for
the four soliton types and different anisotropy ratios.  We have
also obtained a phase diagram showing regions of soliton stability
against collapse and subregions of mobility in terms of the
initial kick-in speed and anisotropy ratio. The phase diagram
shows that 2D solitons become highly mobile for anisotropy ratios
larger than some critical values that depend on the initial
kick-in speed. This fact was then exploited to design tracks
within the 2D waveguide array along which the soliton can be
accelerated and routed. We have calculated the actual waveguide
separations needed to realist the proposed guided trajectories of
2D solitons. \newline
\end{abstract}

\pacs{03.75.Lm; 05.45.Yv; 42.65.Tg, 42.65.Wi}
\maketitle

\section{Introduction.} Discrete solitons appear in many systems 
such as optical waveguide arrays or optical lattices \cite{1,100}. 
Discreteness  introduces new features in the stability and mobility 
of solitons as compared with their continuum counter parts, such as 
mobility threshold \cite{4}, discrete self-trapping \cite{6}, bistability \cite{7},
collisions \cite{7,8}, and the presence of the Peierls-Nabarro (PN)
effective potential  \cite{110,111,112,113}. Existence of
stationary solitons, their mobility, and interaction have been
well-studied \cite{1,100,101,102,1222}.  The discrete nonlinear
Schr$\rm\ddot o$dinger equation, was solved using variational,
perturbative, and numerical approaches  \cite{2,120,121}.
Specifically, the height of the PN potential for highly localised
nonlinear modes was calculated in Ref. \cite{3} and the two
on-site and inter-site stationary states were obtained in Ref.
\cite{5}. The profile of the PN potential has been calculated in
Refs. \cite{3,10} . The potential applications of discrete
solitons in data processing described by the nonlinear
Schr$\rm\ddot o$dinger equations  with various kinds of
nonlinearities, such as unidirectional flow, switching and logic
gates \cite{usapatent}, were ones of the most studied in the field
\cite{1}.

Two-dimensional discrete solitons have also gained considerable
interest especially due to the additional advantages in data
processing applications introduced by the dimensionality
\cite{4nn,14nn,15nnn} and in particular after their experimental
observation in optically induced nonlinear photonic lattices
\cite{obs}. Unlike their continuum counterparts, discrete 2D
solitons are stable against collapse. This is a unique feature
introduced by the discreetness \cite{rev1}. However, 2D discrete
solitons have this natural tendency to collapse but that results
only in narrowing their profile which leads to stronger pinning by
the PN potential and hence low mobility \cite{rev1,10nn,11nn}.
Different setups have thus been considered to explore the
existence, stability, and mobility including waveguides with
modulated nonlinearity \cite{20nn}, 2D solitons in dipolar
condensates \cite{21nn}, rotating waveguide arrays \cite{22nn},
waveguide arrays with defects \cite{23nn} , waveguide arrays with
PT-symmetric couplers \cite{24nn}, and waveguide setups for the
nonlinear Dirac equation \cite{25nn}. The fundamental 2D
stationary discrete solitons were first constructed in
Ref.\cite{15nn} using finite difference numerical method. The
three stationary soliton types found are the so-called
Site-Centered, Bond-Centered, and Hybrid solitons (See Figs.
\ref{fig1}-\ref{fig4} below).  In Ref. \cite{15nnn}, the authors
propose a unique method of routing 2D solitons using `blocker'
high intensity solitons.

In the present work, we consider an anisotropic waveguide array
where the strength of the coupling coefficients in one direction
is larger than in the other, to investigate the role of anisotropy
on the existence, stability, and mobility of the 2D solitons. We
start by looking for the stationary 2D solitons where we found, in
addition to the known soliton types for the isotropic case, that
the hybrid soliton splits into two types with very different
profiles. Investigating the role of anisotropy on the stability of
the 2D solitons, we found a phase diagram of stable solitons in
terms of the strengths of the coupling in the two directions. The
diagram showed a region of stability separated by a sharp border
line from the unstable solitons region. Based on these results, we
show that with anisotropy management along pre-designed tracks the
soliton can be guided to follow the track preserving its integrity
to a large extent. This opens the possibility for all-optical data
processing in two dimensions.

We follow the numerical technique developed by Ref. \cite{15nn} to
find the stationary fundamental solitons in an anisotropic
waveguide array. It turns out that due to the anisotropy, the
hybrid soliton splits into two different soliton which we denote
here as Hybrid-X and Hybrid-Y, as the first is elongated along one
direction and the the second is elongated along the other
direction. For the sake of analytical insight and future
investigations, we perform a variational calculation with a
gaussian trial function and six variational parameters
corresponding to the two components of the soliton position, width, and velocity. The variational calculation accounts for the four
fundamental modes and gives the profile of the PN potential in
terms of the indices of the waveguide array in the two directions.
It is noted here that other trial functions have been used in the
literature such as the hyperbolic secant function and the
kusp-like function for the 1D case \cite{3,5} and the 2D case
\cite{17nn,18nn}, but as argued in \cite{usamapn}, the gaussian
trial function has the advantage of leading to an analytical
profile of the PN potential, which we derive here.  It is then
clearly shown how the anisotropy reduces the PN barrier along one
direction rendering the 2D soliton mobile along that direction.
Both variational and numerical calculations are then used to
generate a phase diagram for the stability and mobility of the 2D
solitons in terms of the anisotropy.

Having determined the mobility region in the phase diagram, we
design tracks within the 2D waveguide array along which the
coupling coefficients satisfy the anisotropy required for mobility
and modulated in their strength such that they are site-dependent.
This is equivalent to an effective potential \cite{andrey, usama}
which we choose to be a linear potential that leads to soliton
motion with constant acceleration. With a track composed of three
segments perpendicular to each other, the soliton is then guided
along these tracks preserving its integrity. Based on an
experimental calibration of the strength of couplings in terms of
separation \cite{exptcouplings}, we have calculated the waveguides
separations in $\rm\mu$m that are expected to result in such a
guided trajectory.

The rest of the paper is organized as follows. In section
\ref{stationary_sec} we use  numerical and variational
calculations to construct the four stationary fundamental
anisotropic 2D solitons. We also investigate their stability and
calculate their stability phase diagram in terms of the
anisotropy. In section \ref{mob_sec}, we investigate the mobility
of the anisotropic 2D solitons and calculate a phase diagram that
shows regions of stable mobile 2D solitons in terms of anisotropy.
Then in section \ref{manage_sec}, we employ this fact to design
tracks where solitons are accelerated and routed. Finally, we end
with  summarizing our main conclusions and discussing some future
follow ups in section \ref{conc_sec}.

\section{Anisotropic 2D solitons and PN Potential}\label{stationary_sec}
In this section, we construct the four stationary fundamental anisotropic 2D solitons using both numerical procedure, in section \ref{num_sec}, and variational approach in section \ref{var_sub}. With a gaussian trial function we derive analytical expressions for the 2D PN potential surface. We calculate in this section the PN barrier depths for the four types of solitons using two trial functions, namely the gaussian and the kusp-like exponential function. Finally we calculate in section \ref{stab_sec} a phase diagram for the stability of the 2D anisotropic solitons against collapse in terms of the anisotropy.

\section{Model Equation and Numerical Procedure}\label{num_sec}
The scaled 2D discrete nonlinear Schrödinger equation
describing propagation of solitons in anisotropic waveguide
arrays can be written, in a straightforward generalization to the anisotropic case, as \cite{40scale}
\begin{align}
i&\frac{\partial}{\partial_t}\Psi_{i,j}+d_x\,\Psi_{i-1,j}+d_x\,\Psi_{i+1,j}+d_y\Psi_{i,j-1}\nonumber \\
&+d_y\Psi_{i,j+1}-2(d_x+d_y)\Psi_{i,j}+\gamma\left|\Psi_{i,j}\right|^2\Psi_{i,j}=0,
\label{eq1}
\end{align}
where, $\Psi_{i,j}$ is the field variable at the site $(i,j)$, $\gamma$ is the strength of the nonlinearity which is assumed to be positive in order to support bright solitons,  and $d_x$ and $d_y$ are the coupling coefficients between waveguides in the horizontal and vertical directions, respectively.  Trivially, for $d_x=d_y$ the isotropic case is retrieved. Discrete nonlinear Schr\"odinger equations are derived from a tight-binding model where the coupling coefficients correspond to the evanescent interaction between the modes in neighboring waveguides. The 2D waveguide array can thus be set up such that the coupling between the waveguides along one direction is stronger than the other. This can be performed, for instance, by setting the waveguides along, say the horizontal direction, closer to each other than for the perpendicular waveguides along the vertical direction, which results in coupling coefficients along the horizontal direction being larger than in the vertical direction. Such an anisotropic setup is indeed described by Eq. (\ref{eq1}).

The isotropic version of Eq. (\ref{eq1}) supports three stationary 2D soliton types, known in the literature as the Site-Centered, Bond-Centered, and Hybrid solitons. Some references use though other names. The purpose of this section is to investigate the role of the anisotropy on the existence and profile of the stationary modes which requires solving Eq. (\ref{eq1}) numerically. For a numerical procedure, we employ here a slightly modified version of the finite difference method developed by Ref. \cite{15nn}. We assume an $L$ $\times$ $L$ dimensional square lattice. The initial condition is given in matrix form $\left[H\right]$ of the following type, namely
\begin{eqnarray}
\left[H\right]_{n,m}&=&2\,d_x+2\,d_y-\gamma\lvert\Psi_{n,m}\rvert^2,\\
\left[H\right]_{n+1,m}&=&\left[H\right]_{n-1,m}=-d_x,\\
\left[H\right]_{n,m+1}&=&\left[H\right]_{n,m-1}=-d_y,
\end{eqnarray}
where, $n = i + (l-1)j$ and $m = j + (l-1)i$,  $l=1,2,....L$ for the square lattices of size $L \times L$. Solving the linear eigenvalue problem refines
the prediction of $\Psi_{i,j}$ as the eigenfunction corresponding
to the most negative eigenvalue. This procedure is repeated until the desired precision is reached.

\begin{figure}
\includegraphics[width=1.0\linewidth]{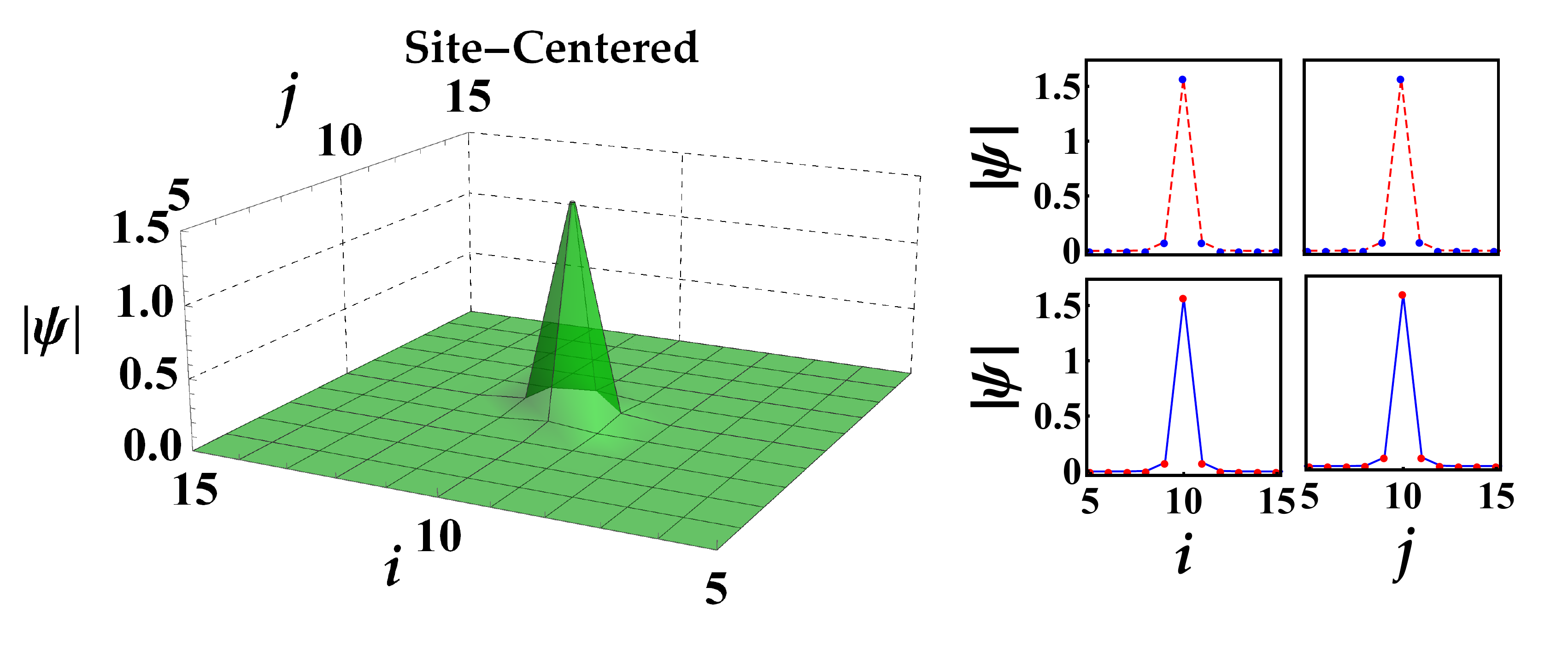}
\caption{Isotropic Site-Centered (SC) soliton. Obtained from the numerical solution of Eq. (\ref{eq1}). We use the trial function given by Eq. \eqref{SC} and the parameters $L=20,\,P=2.5,\,\gamma=4$ with $d_x=d_y=0.2$.  The plots on the right show the two cross-section profiles (points). The lines correspond to the variational calculation using a gaussian trial function, Eq. (\ref{gaussian}), (red dashed line) and kusp-like exponential trial function, Eq. (\ref{expo}), (blue solid line). }\label{fig1}
\end{figure}

\begin{figure}
    \includegraphics[width=1.0\linewidth]{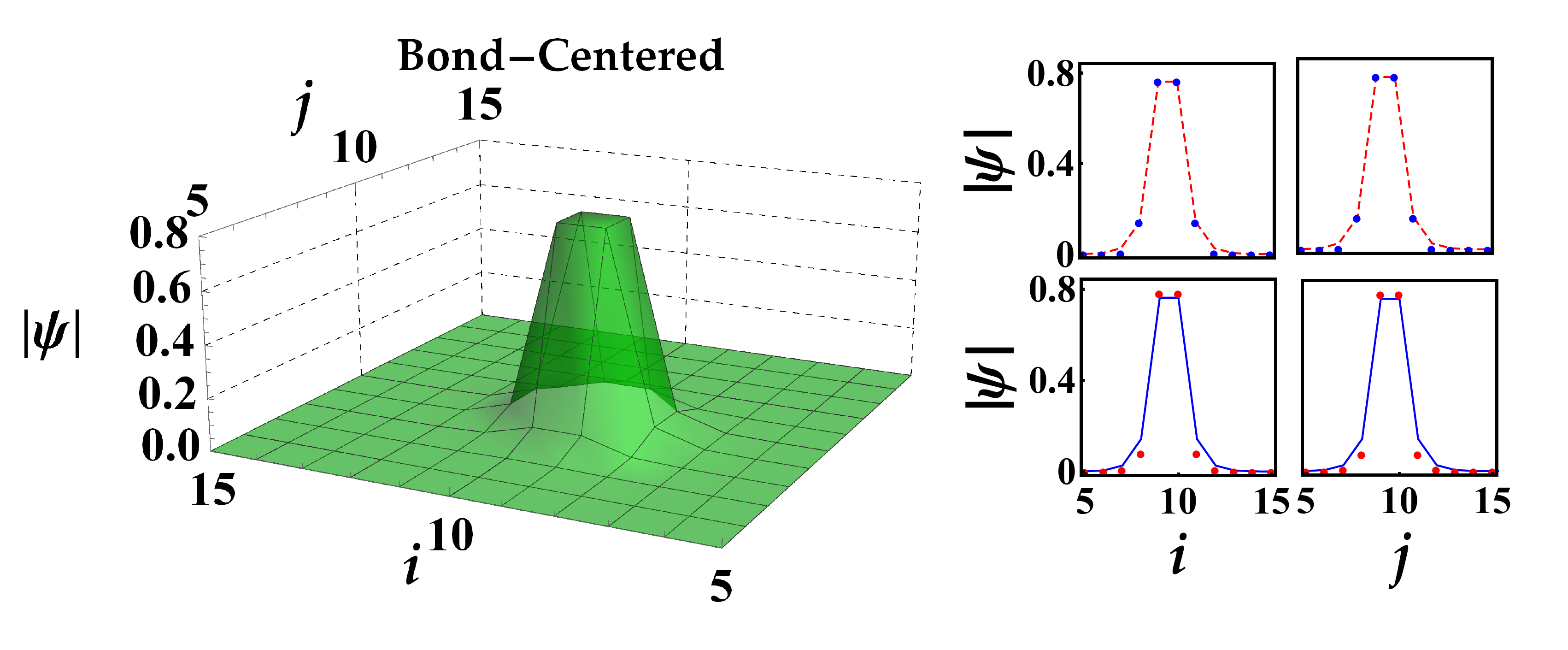}
    \caption{Isotropic Bond-Centered (BC) soliton. Trial function \eqref{BC} and the parameters of Fig.\ref{fig1} are used. }\label{fig2}
\end{figure}

\begin{figure}
    \includegraphics[width=1.0\linewidth]{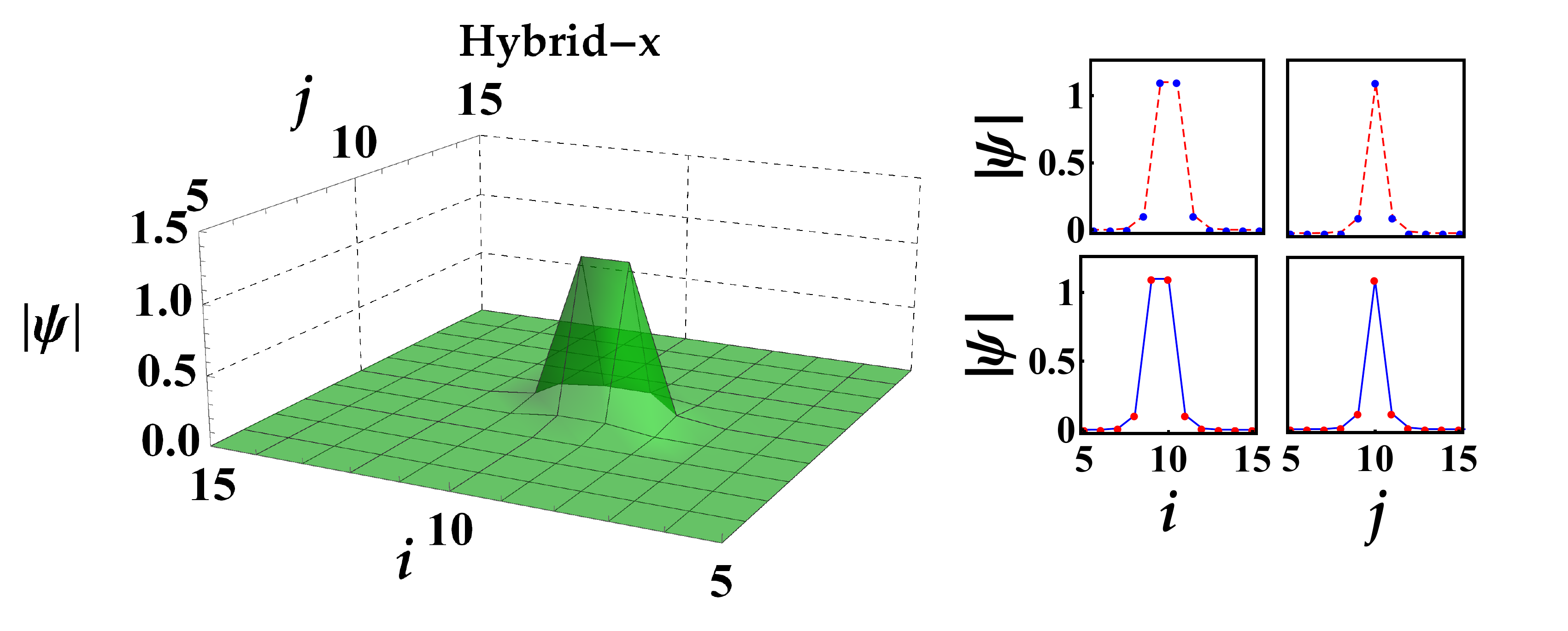}
    \caption{Isotropic Hybrid-X (HX). Trial function \eqref{HX} and the parameters of Fig.\ref{fig1} are used.}\label{fig3}
\end{figure}

\begin{figure}
    \includegraphics[width=1.0\linewidth]{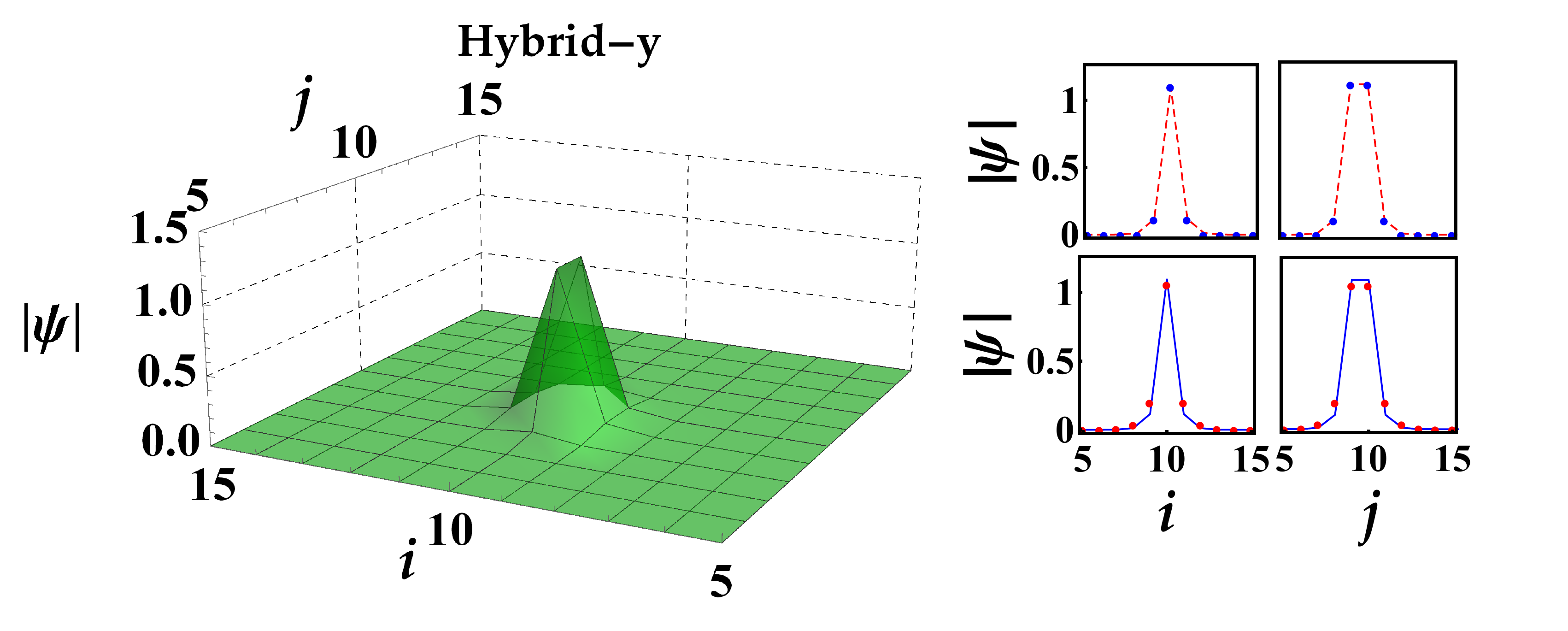}
    \caption{Isotropic Hybrid-Y(HY)  soliton. Trial function \eqref{HY} and the parameters of Fig.\ref{fig1} are used. }\label{fig4}
\end{figure}

\begin{figure}
    \includegraphics[width=1.0\linewidth]{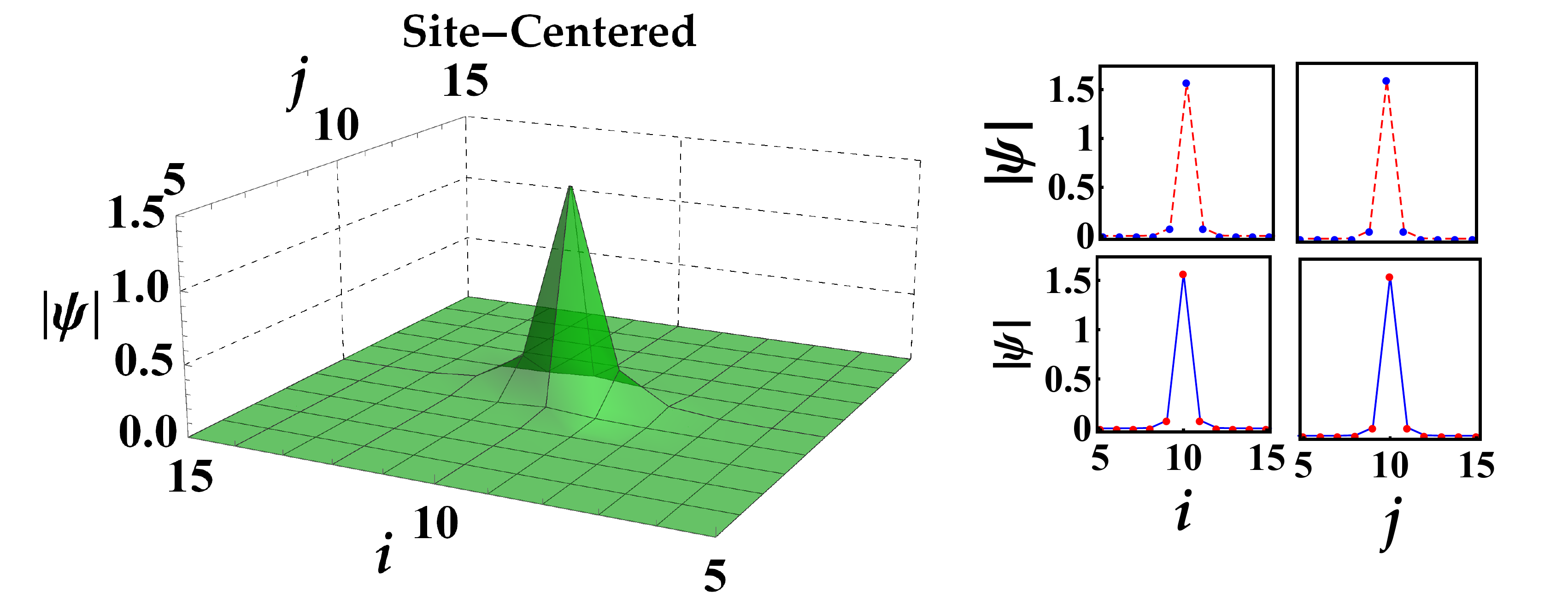}
    \caption{Anisotropic Site-Centered soliton. Trial function \eqref{SC} and the parameters of Fig.\ref{fig1} are used, but with $d_x=1.5,\,d_y=0.2$. }\label{fig5}
\end{figure}

\begin{figure}
    \includegraphics[width=1.0\linewidth]{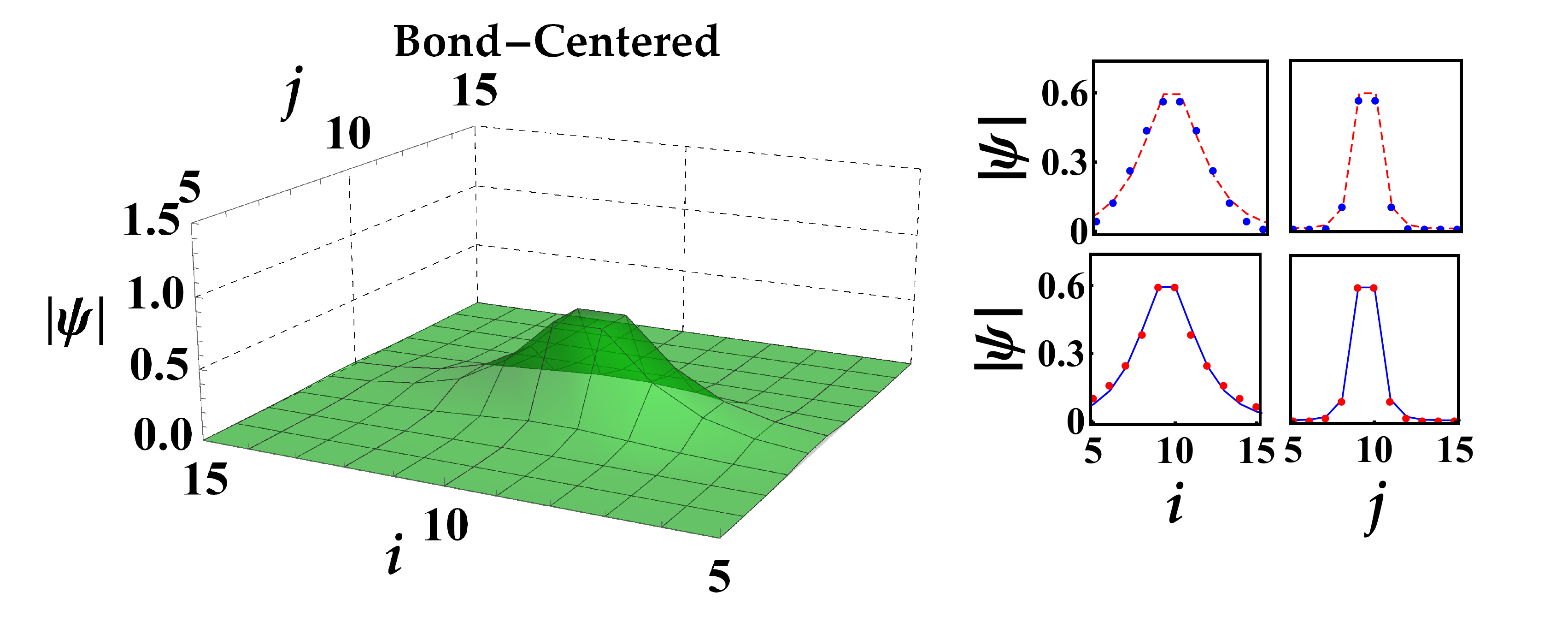}
    \caption{Anisotropic Bond-Centered soliton. Trial function \eqref{BC} and the parameters of Fig.\ref{fig5} are used.  }\label{fig6}
\end{figure}

\begin{figure}
    \includegraphics[width=1.0\linewidth]{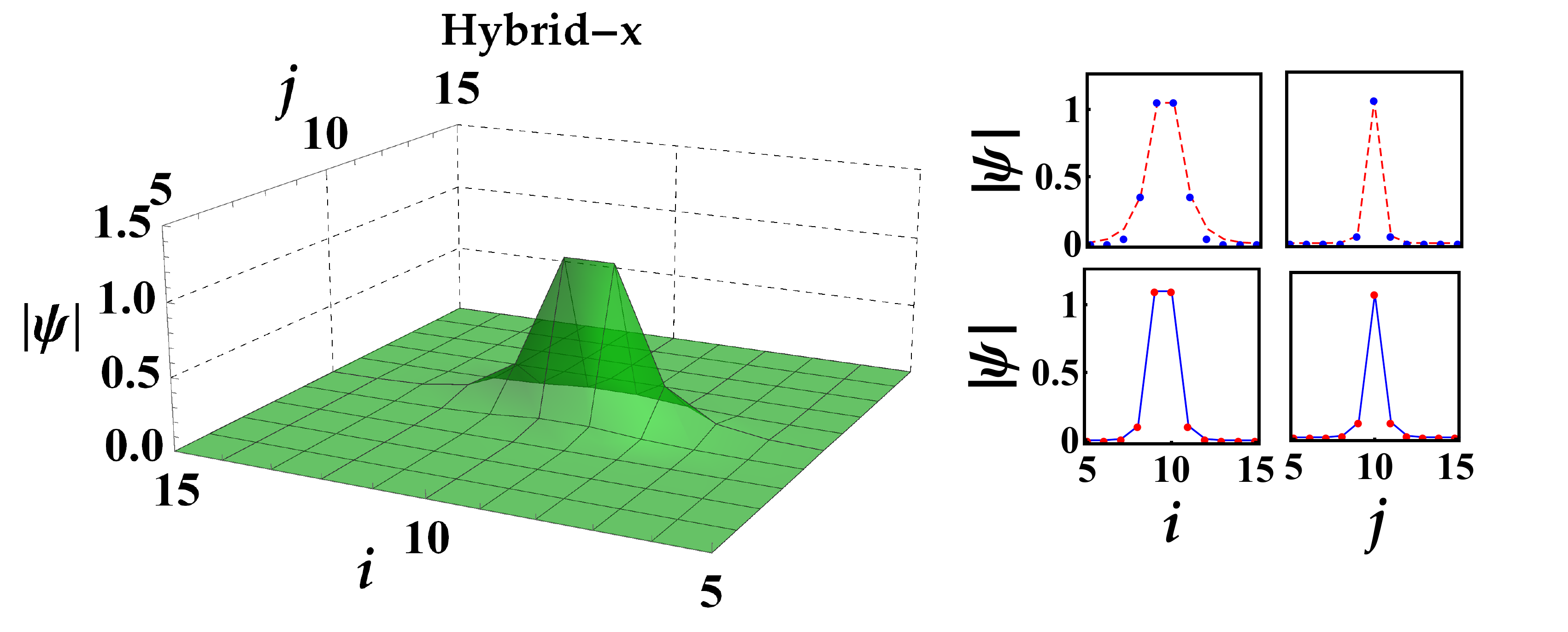}
    \caption{Anisotropic Hybrid-X soliton. Trial function \eqref{HX} and the parameters of Fig.\ref{fig5} are used.}\label{fig7}
\end{figure}

\begin{figure}
    \includegraphics[width=1.0\linewidth]{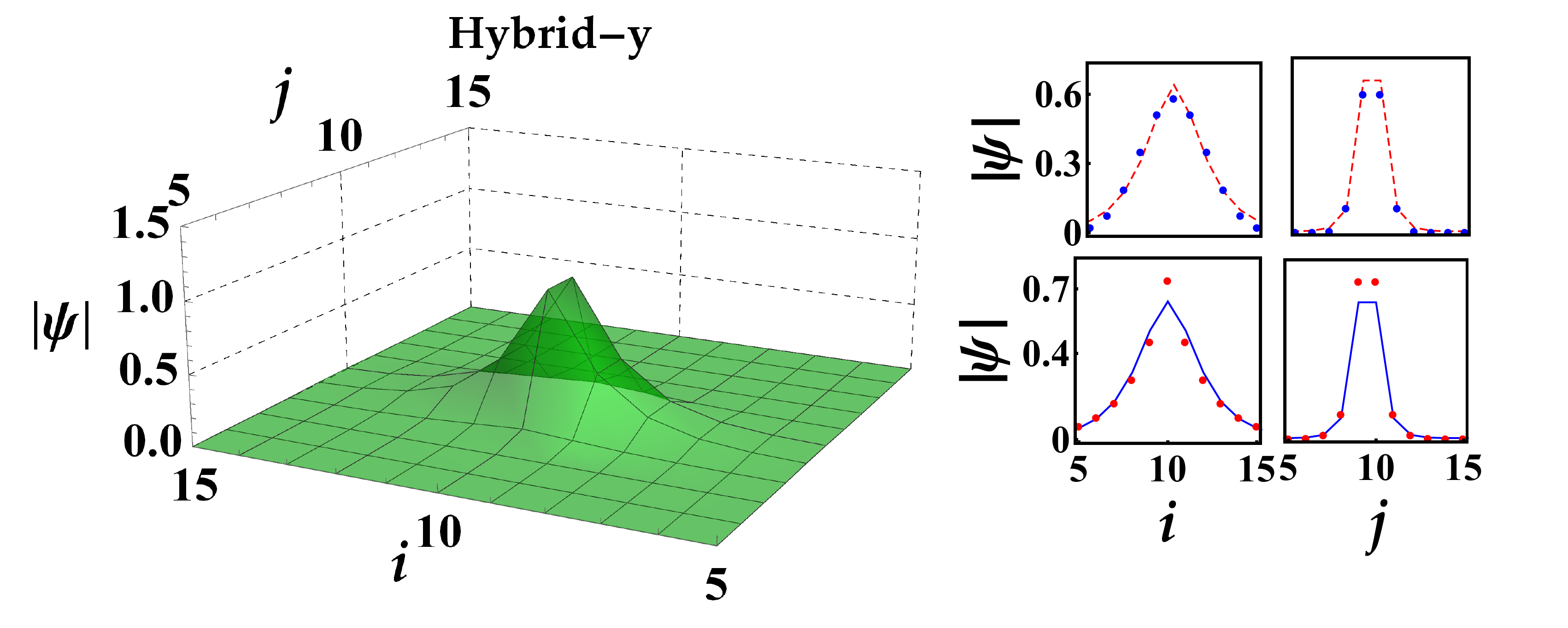}
    \caption{Anisotropic Hybrid-Y soliton. Trial function \eqref{HY} and the parameters of Fig.\ref{fig5} are used. }\label{fig8}
\end{figure}

Using four different trial functions given by
\begin{equation}
\Psi_{i,j}^{SC}=A\, e^{-\left(\lvert i-{L}/{2}\rvert+\lvert j-{L}/{2}\rvert\right)},\label{SC}
\end{equation}
\begin{equation}
\Psi_{i,j}^{BC}=A\, e^{-\left(\lvert i-{L}/{2}+1/2\rvert+\lvert j-{L}/{2}+1/2\rvert\right)},\label{BC}
\end{equation}
\begin{equation}
\Psi_{i,j}^{HX}=A\, e^{-\left(\lvert i-{L}/{2}+1/2\rvert+\lvert j-{L}/{2}\rvert\right)},\label{HX}
\end{equation}
and
\begin{equation}
\Psi_{i,j}^{HY}=A\, e^{-\left(\lvert i-{L}/{2}\rvert+\lvert j-{L}/{2}+1/2\rvert\right)}\label{HY},
\end{equation}
to solve the model given by Eq. (\ref{eq1}) we have found four types of solitons, as shown in Figs.\ref{fig1}-\ref{fig8} which we denote as Site-Centered (SC), Bond-Centered (BC), Hybrid-X (HX) and Hybrid-Y (HY).

For the purposes of checking our numerical calculation, we have generated first the three 2D soliton types of the isotropic case, as shown by Figs.\ref{fig1}-\ref{fig4}. It is clear that the two hybrid solitons, in Figs. \ref{fig3} and \ref{fig4}, are in this case equivalent; one is merely the $90^{\rm o}$-rotation of the other. Therefore, these are considered in the literature as one type and denoted as just Hybrid soliton. On the other hand, the two hybrid solitons for the anisotropic case, see Figs. \ref{fig7} and \ref{fig8}, are different and not related to each other by a rotation. Hence the two names HX and HY. The effect of anisotropy on the other two solitons, SC and BC, is a mere elongation in the direction of larger coupling.

The general feature of anisotropy elongating the solitons along one direction as compared with the isotropic case has the effect of enhancing the mobility of the soliton in that direction, as we will see below.  This will be confirmed in the next section where we calculate and plot the PN potential and show that the PN barrier decreases in the direction of larger coupling.

\section{Variational Approach and the PN Potential}\label{var_sub}
In this section, we use a variational calculation to account for the four stationary solitons found numerically in the previous section and then derive an analytical expression for the PN potential in terms of the two waveguide indices which will enable us to calculate the PN barrier in any direction and for any of the four soliton types. This will provide an insight on the role of anisotropy in enhancing the mobility of the 2D solitons. Furthermore, we will be able to calculate the phase diagram for soliton stability against collapse.

The Lagrangian corresponding to the  above 2D discrete nonlinear
Schr\"odinger equation, Eq. (\ref{eq1}), is written as

\begin{align}
L=&\sum_{i=-\infty}^{\infty}\sum_{j=-\infty}^{\infty}\left[\frac{i}{2}\left(\Psi_{i,j}\frac{\partial}{\partial t}\Psi_{i,j}^{\ast}-\Psi_{i,j}^{\ast}\frac{\partial}{\partial t}\Psi_{i,j}\right) \right. \nonumber \\ &+ \left. \Psi_{i,j}^{\ast}\left(d_x\Psi_{i-1,j}+d_x\Psi_{i+1,j}+d_y\Psi_{i,j-1}+d_y\Psi_{i,j+1} \right. \right. \nonumber \\ &- \left. \left. (d_x+d_y)\Psi_{i,j}\right)+\frac{1}{2}\gamma\left|\Psi_{i,j}\right|^4 \right]
\end{align}

where, the dispersion and nonlinear terms define the energy functional
\begin{align}
E=&-\sum_{i=-\infty}^{\infty}\sum_{j=-\infty}^{\infty}\left[
\Psi_{i,j}^{\ast}\left(d_x\,\Psi_{i-1,j}+d_x\,\Psi_{i+1,j}+d_y\Psi_{i,j-1}\right. \right. \nonumber \\ & \left.\left. +d_y\Psi_{i,j+1}-2(d_x+d_y)\Psi_{i,j}\right)+\frac{1}{2}\,\gamma\left|\Psi_{i,j}\right|^4\right].
\label{efunc}
\end{align}

As for the trial function, three options are available, namely the kusp-like exponential function \cite{120,5}, the hyperbolic secant function \cite{3}, and the gaussian function \cite{usamapn}. With the kusp-like exponential trial function analytic expression for the largrangian can be obtained in a compact form. However, this requires knowledge of whether the soliton is peaked at a site or between two sites. Thus, as argued in Ref. \cite{usamapn}, the dynamics of center-of-mass of a soliton that travels across the sites cannot
be obtained with this trial function. For the hyperbolic secant trial function, the sums in the lagrangian can not be performed in a compact form and only asymptotic expressions can be obtained in the large soliton width limit. The gaussian trial function has been used by Ref. \cite{usamapn} where it was shown that the langrangian can be obtained in a compact analytic form without a priori assumptions about the location of the peak of the soliton, which lead to an account of the soliton motion across the sites and to a profile of the PN potential in terms of the soliton's location. For this reason, we use here the gaussian trial function to derive the 2D solitons' profile and PN potential. We will also re-calculate the solitons' profile and PN barriers using the kusp-like trial function in the next section for the purpose of comparing the two trial functions with each other and with the exact numerical solution.

\begin{figure}
    \includegraphics[width=1.0\linewidth]{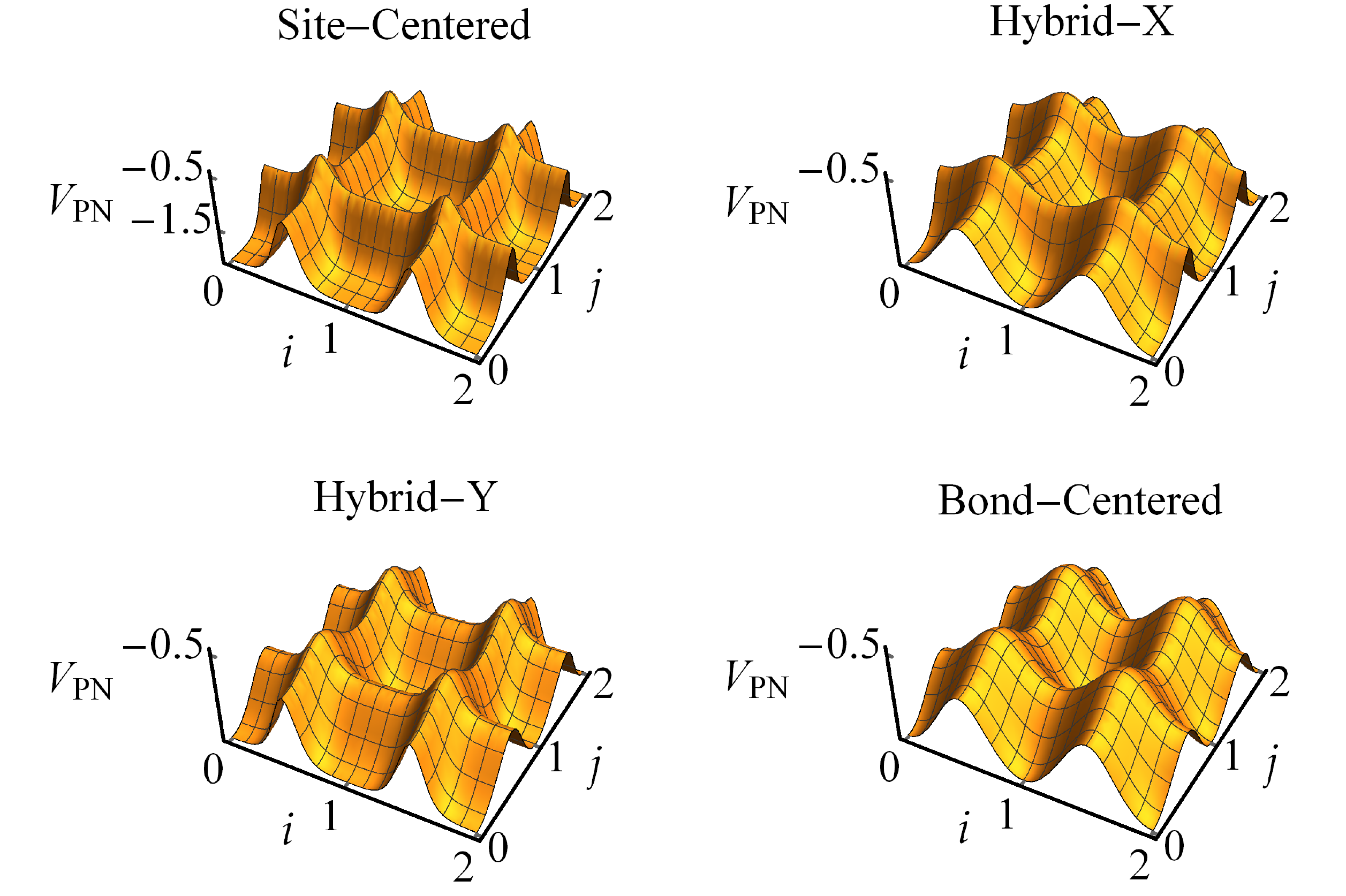}
    \caption{The PN potential for the isotropic case for the same choice of parameters used in Figs. \ref{fig1}-\ref{fig4}, respectively.}\label{fig9}
\end{figure}

The  gaussian trial function reads
\begin{equation}
\psi_{i,j}^{g}=A \,e^{-\frac{(i-n_1)^2}{\eta_1^2}-\frac{(j-n_2)^2}{\eta_2^2}+\mathit{i}v_1(i-n_1)+\mathit{i}v_2(j-n_2)},\label{gaussian}
\end{equation}
where $A$ is the normalization constant, and the coordinates of the peak position, $n_{1,2}$,  the widths of the soliton in the horizontal and vertical directions, $\eta_{1,2}$, and the group velocity components in the two directions, $v_1,2$, are six variational parameters. The first step is to
normalise the trial function given by Eqs.\eqref{gaussian} to the
constant power $P$
\begin{equation}
P=\sum_{i=-\infty}^{\infty}\sum_{j=-\infty}^{\infty}|\Psi_{i,j}^{g}|^{2}
\end{equation}
which gives $A$ in terms of the Elliptical function, $\vartheta_3\left(x\right)$,
\begin{equation}
A=\frac{\sqrt{P}}{\sqrt{\frac{1}{2}\pi\eta_1\eta_2
\vartheta_3\left(-n_1\pi
e^{-\frac{1}{2}\pi^2\eta_1^2}\right)\vartheta_3\left(-n_2\pi\,e^{-\frac{1}{2}\pi^2
\eta_2^2}\right)}}
\end{equation}
The normalized trial function is then used to calculate the energy functional
\begin{align}
E&[v_1,v_2,n_1,n_2,\eta_1,\eta_2]\nonumber \\
=&-\frac{P^2 \gamma \vartheta _3 \left(-n_1\pi,
e^{-\frac{1}{4}\pi^2\eta_1^2}\right) \vartheta_3\left(-n_2\pi,
e^{-\frac{1}{4} \pi^2 \eta_2^2}\right)}{2\pi\eta_1\eta_2
\vartheta_3\left(-n_1\pi,e^{-\frac{1}{2}\pi^2 \eta_1^2}\right)^2
\vartheta_3\left(-n_2 \pi,e^{-\frac{1}{2}\pi^2\eta_2^2}\right)^2}\nonumber\\
&-2\,P\, \left(-(d_x+d_y)+ d_x\,\cos(v_1)E_1 +d_y\,\cos(v_2)E_2\right).\nonumber\\
\end{align}
where
\begin{eqnarray}
E_1=&&\frac{e^{-\frac{1}{2 \eta_1^2}}
\vartheta_3\left[-\frac{1}{2}(1+2n_1)\pi,e^{-\frac{1}{2} \pi^2
\eta_1^2}\right]}{\vartheta _3\left(-n_1 \pi, e^{-\frac{1}{4}
\pi^2 \eta_1^2}\right)},\\ E_2&=&\frac{e^{-\frac{1}{2 \eta_2^2}}
\vartheta_3\left[-\frac{1}{2}(1+2n_2)\pi,e^{-\frac{1}{2}\pi^2
\eta_2^2}\right]}{\vartheta_3\left(-n_2 \pi, e^{-\frac{1}{4} \pi^2
\eta_2^2}\right)}.
\end{eqnarray}
Stationary solitons are obtained by minimizing the energy functional with
respect to the parameters $v_1,v_2,\eta_1,\eta_2,n_1,n_2$. By inspection, it is found that the energy functional is minimum for stationary solitons, $v_1,\, v_2=0$, for all values of the other variational parameters. Setting this condition in the energy functional gives the PN potential
\begin{align}
V_{\rm PN}&(n_1,n_2,\eta_1,\eta_2)\nonumber \\ 
=& E[0,0,n_1,n_2,\eta_1,\eta_2] \nonumber \\ 
=&-\frac{P^2 \gamma \vartheta _3 \left(-n_1\pi,
	e^{-\frac{1}{4}\pi^2\eta_1^2}\right) \vartheta_3\left(-n_2\pi,
	e^{-\frac{1}{4} \pi^2 \eta_2^2}\right)}{2\pi\eta_1\eta_2
	\vartheta_3\left(-n_1\pi,e^{-\frac{1}{2}\pi^2 \eta_1^2}\right)^2
	\vartheta_3\left(-n_2 \pi,e^{-\frac{1}{2}\pi^2\eta_2^2}\right)^2}\nonumber\\
&-2\,P\, \left(-(d_x+d_y)+ d_x\,E_1 +d_y\,E_2\right).\label{PNP}
\end{align}
It is noticed that $V_{\rm PN}$ is periodic in both $n_1$ and $n_2$ with periods equal $n_1\pi$ and $n_2\pi$, respectively. This leads to four stationary points in $V_{\rm PN}$ in terms of $n_1$ and $n_2$; determined by $n_1$ and $n_2$ being integer or half integer. Each one of these four cases will correspond to a stationary 2D soliton, which will be identified below as the ones obtained above numerically, namely the SC, BC, HX, and HY solitons. The equilibrium widths of these solitons are obtained by minimizing $V_{\rm PN}$ with respect to $\eta_1$ and $\eta_2$. For instance, when $n_1$ and $n_2$ are both integers, which without loss of generality can be taken as $n_1=n_2=0$,  the 2D soliton will be the Site-Centered soliton and the PN potential for this specific case takes the form
\begin{equation}
V_{\rm PN}^{\rm SC}(\eta_1,\eta_2)=V_{\rm PN}(0,0,\eta_1,\eta_2)
   .
\end{equation}
The equilibrium width of this soliton type will be given by
\begin{equation}
\frac{\partial V^{\rm SC}_{\rm PN}(\eta_1,\eta_2)}{\partial \eta_1}\lvert_{\eta_1=\eta_{1eq}^{\rm SC}}\,\,=0,\hspace{1cm}\frac{\partial V_{\rm PN}^{\rm SC}(\eta_1,\eta_2)}{\partial \eta_2}\lvert_{\eta_2=\eta_{2eq}^{\rm SC}}\,\,=0,
\end{equation}
where $\eta_{1eq}^{\rm SC}$ and $\eta_{2eq}^{\rm SC}$ are the equilibrium widths of the SC soliton.
Substituting back the equilibrium widths in $V_{\rm PN}(n_1,n_2,\eta_1,\eta_2)$, we obtain the PN potential for the SC soliton in terms of $n_1$ and $n_2$, namely
\begin{equation}
V_{\rm PN}^{\rm SC}(n_1,n_2)=V_{\rm PN}(n_1,n_2,\eta_{1eq}^{\rm SC},\eta_{2eq}^{\rm SC}),
\end{equation}
which is plotted in Fig. \ref{fig9}. The SC soliton is peaked at the minimum of the PN potential. The depth of the PN potential at the soliton peak is given by $E_{\rm min}^{\rm SC}=V_{\rm PN}^{\rm SC}(\eta_{1eq},\eta_{2eq})$. The equilibrium soliton profiles can be obtained by substituting the equilibrium widths in the variational function, Eq. (\ref{gaussian}). The variational soliton profiles across the $i$th and $j$th cross sections are then plotted in Fig. \ref{fig1} where excellent agreement with the numerical profiles is observed.
The PN potentials for the other three soliton types are similarly given by
\begin{eqnarray}
V_{\rm PN}^{\rm BC}(\eta_1,\eta_2)&=&V_{\rm PN}\left(\frac{1}{2},\frac{1}{2},\eta_1,\eta_2\right)
   ,
\end{eqnarray}
\begin{equation}
V_{\rm PN}^{\rm HX}(\eta_1,\eta_2)=V_{\rm PN}\left(\frac{1}{2},0,\eta_1,\eta_2\right)
   ,
\end{equation}
\begin{equation}
V_{\rm PN}^{\rm HY}(\eta_1,\eta_2)=V_{\rm PN}\left(0,\frac{1}{2},\eta_1,\eta_2\right)
  .
\end{equation}
Minimising these potentials with respect to $\eta_1$ and $\eta_2$ and then substituting back in $V_{\rm PN}(n_1,n_2,\eta_1,\eta_2)$, we obtain the PN potential for each soliton type
\begin{eqnarray}
V_{\rm PN}^{\rm BC}(n_1,n_2)&=&V_{\rm PN}(n_1,n_2,\eta_{1eq}^{\rm BC},\eta_{2eq}^{\rm BC}),\\
V_{\rm PN}^{\rm HX}(n_1,n_2)&=&V_{\rm PN}(n_1,n_2,\eta_{1eq}^{\rm HX},\eta_{2eq}^{\rm HX}),\\
V_{\rm PN}^{\rm HY}(n_1,n_2)&=&V_{\rm PN}(n_1,n_2,\eta_{1eq}^{\rm HY},\eta_{2eq}^{\rm HY}).
\end{eqnarray}
The four PN potentials are plotted in Fig. \ref{fig9} and the variational profiles are plotted in Figs. \ref{fig1}-\ref{fig4} together with the numerical profiles.

As can be seen in Fig. \ref{fig9}, the primitive cell of the periodic PN potential is bounded by two barriers parallel to the $i$th direction and two barriers parallel to the $j$th direction. The SC soliton is peaked at the centre of the cell, the BC soliton has an equal amplitude at the centres of 4 nearest neighbour cells, the HX soliton has an equal amplitude at the centres of two neighboring cells aligned along the $i$th direction,  and the HY soliton has an equal amplitude at the centres of two neighboring cells aligned along the $j$th direction. The depth of the PN potential at the point where the soliton is peaked is a characteristic value for the potential. Therefore, the PN potential has four characteristic energy barriers defined by
\begin{eqnarray}
E_{\rm min}^{\rm SC}&=&V_{\rm PN}\left(0,0,\eta_{1eq}^{\rm SC},\eta_{2eq}^{\rm SC}\right),\\
E_{\rm min}^{\rm BC}&=&V_{\rm PN}\left(\frac{1}{2},\frac{1}{2},\eta_{1eq}^{\rm BC},\eta_{2eq}^{\rm BC}\right),\\
E_{\rm min}^{\rm HX}&=&V_{\rm PN}\left(\frac{1}{2},0,\eta_{1eq}^{\rm HX},\eta_{2eq}^{\rm HX}\right),\\
E_{\rm min}^{\rm HY}&=&V_{\rm PN}\left(0,\frac{1}{2},\eta_{1eq}^{\rm HY},\eta_{2eq}^{\rm HY}\right),
\end{eqnarray}
where $E_{\rm min}^{\rm SC}$, $E_{\rm min}^{\rm BC}$, $E_{\rm min}^{\rm HX}$, and $E_{\rm min}^{\rm HY}$ are the PN barrier depths at the centre of the SC, BC, HX, and HY solitons, respectively. In Table \ref{table1}, we give an example with the specific case of $d_x=d_y=0.2$, where we calculate the variational equilibrium widths of the four soliton types and their PN barrier. It is noticed that the SC soliton has the largest barrier depth and therefore is the most pinned soliton type. On the other hand, the BC soliton is the most mobile soliton since it has the lowest PN barrier depth. Due to the isotropic symmetry, the barrier depths of the HX and HY solitons are equal and the $(\eta_{1eq},\eta_{2eq})$ of the HX soliton are equal to  $(\eta_{2eq},\eta_{1eq})$ of the HY soliton, respectively, which means that the HX and HY solitons are equivalent when one is rotated by $90^{\rm o}$ with respect to the other.

\begin{table*}
\caption{Variational soliton equilibrium widths and energy for the isotropic case with\,\, $d_{\textit{x}}=d_{\textit{y}}=0.2$.}
\centering
\begin{tabular}{|c|c|c|c|c|c|}
\hline Type of Solution\,\,& $n_1$ & $n_2$  & $\eta_{1eq}$ & $\eta_{2eq}$ & $E_{\rm min}$  \\
\hline
Site-Centered  & Integer\, & Integer\, & 0.505721 & 0.505721 & -2.108    \\
Bond-Centered  & Half-Int\,& Half-Int\,& 0.879461 & 0.879461 & -0.439898    \\
Hybrid-X       & Half-Int\,& Integer\, & 0.78443  & 0.557776 & -0.961864    \\
Hybrid-Y       & Integer\, & Half-Int\,& 0.557776 & 0.78443  & -0.961864    \\
\hline
\end{tabular}
\label{table1}
\end{table*}

\begin{table*}
\caption{Variational soliton equilibrium widths and energy for the anisotropic case with \,\, $d_{\textit{x}}$=1.5,
$d_{\textit{y}}$=0.2.} \centering
\begin{tabular}{|c|c|c|c|c|c|}
\hline Type of Solution\,\,& $n_1$ & $n_2$  & $\eta_{1eq}$ & $\eta_{2eq}$ & $E_{\rm min}$  \\
\hline
Site-Centered  & Integer\, & Integer\, & 0.731183 & 0.51165  & -1.02698    \\
Bond-Centered  & Half-Int\,& Half-Int\,& 1.92085  & 0.975147 & -0.0889755   \\
Hybrid-X       & Half-Int\,& Integer\, & 1.20644  & 0.568304 & -0.484191    \\
Hybrid-Y       & Integer\, & Half-Int\,& 0.991607 & 0.838762 & -0.0916029    \\
\hline
\end{tabular}
\label{table2}
\end{table*}

Having established confidence in the variational calculation by accounting for the known 2D stationary solitons, their accurate profiles and widths, and their PN potential profiles and barriers, we move now to use the variational calculation in investigating the effect of anisotropy on all of these quantities and properties of 2D solitons. We start by the numerical solution for the anisotropic case of $d_x=1.5$ and $d_y=0.2$. The profiles of the four soliton types are shown in Figs. \ref{fig5}-\ref{fig8}, with new features that have been discussed in the previous section. Most importantly, the symmetry between the HX and HY solitons is now broken and they are treated as two different types. The variational calculation for such an anisotropic case still gives an accurate account for the soliton widths and profiles in comparison with the numerical values, as shown on the right panels of these figures.

\begin{figure}
	\includegraphics[width=1.0\linewidth]{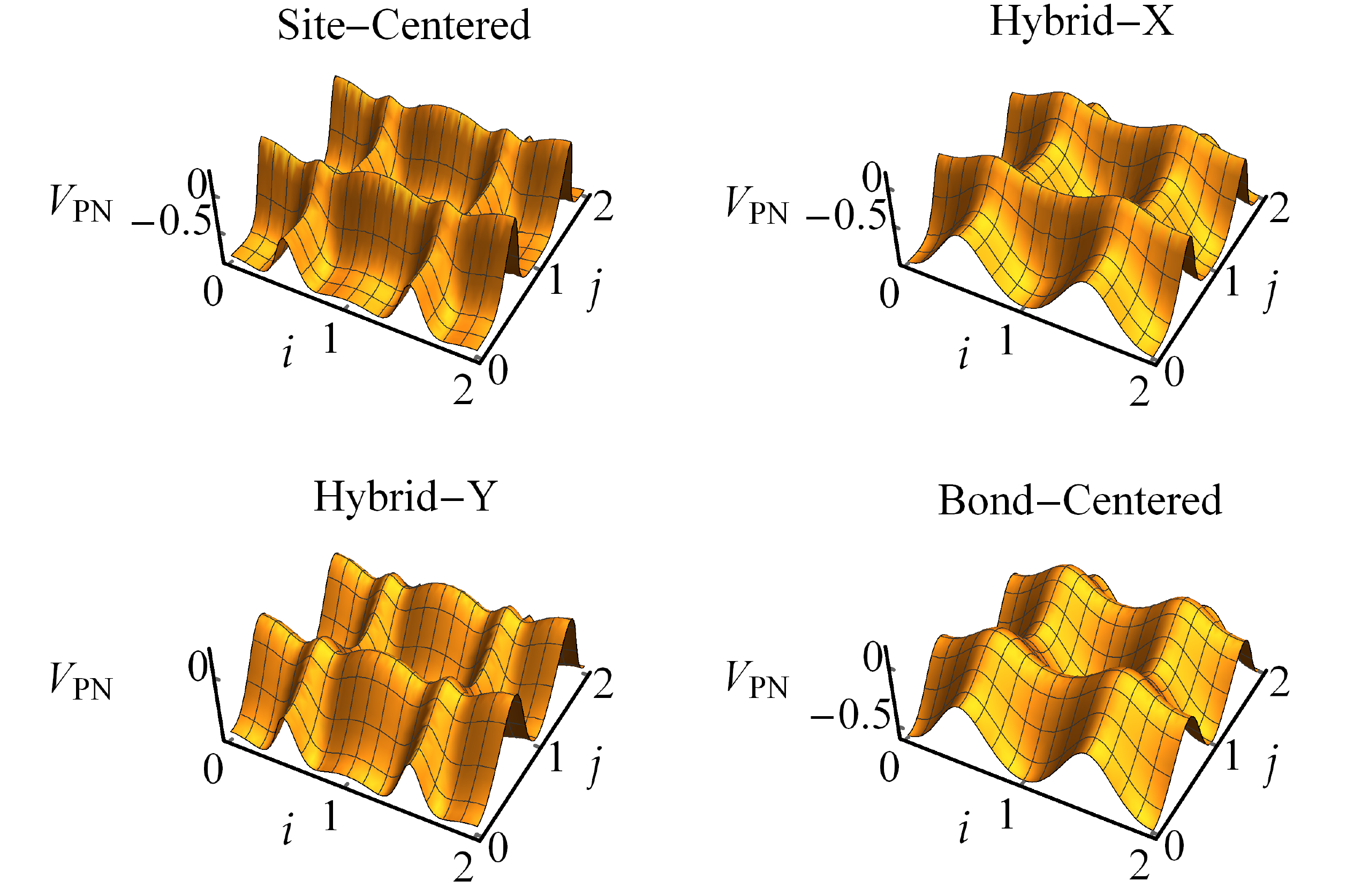}
    \caption{The PN potential for the anisotropic solitons for the same choice of parameters used in Fig. \ref{fig5}-\ref{fig8}.}\label{fig10}
\end{figure}

The mobility of the soliton is determined by the height of the PN potential. By comparing Fig. \ref{fig9} with Fig. \ref{fig10} for the SC soliton, we observe that for the anisotropic case ($d_x>d_y$), the height of the PN barrier in the $i$th direction becomes less than that in the $j$th direction, which means that due to the anisotropy, the mobility of the SC soliton will be enhanced in the direction of the larger coupling. A similar conclusion can be drawn for the other soliton types. To quantify the comparison, we have re-calculated in Table \ref{table2} the solitons' equilibrium widths and PN barrier depths for the anisotropic case. In Table \ref{table1} we used $d_x=d_y=0.2$ and in Table \ref{table2} we used $d_x=1.5$ and $d_y=0.2$, therefore the absolute values of PN barriers should not be compared directly; the total energy is different for the two cases. Instead, we compare the PN barrier heights relative to a reference, which we take the PN height for the SC soliton. For the isotropic case, the PN barrier for the HX relative to the SC soliton is $-0.962/(-2.101)=0.456$ while for the anisotropic case, the ratio is $-0.484/(-1.027)=0.471$, which is slightly more. For the HY soliton, the barriers ratio for the isotropic case is the same as for HX, namely 0.456, while for the anisotropic case it equals 0.089. Keeping in mind that the PN barrier for the HX soliton is the depth of the PN potential at the middle of the horizontal junction in the cell and the PN barrier for the HY soliton is the depth of the PN potential at the middle of the vertical junction in the cell, we conclude that the PN barrier depth of the junction parallel to the $i$th direction drops from 0.96 to 0.09 while the barrier depth for the junction parallel to the $j$th direction drops from 0.96 to 0.48 as a result of an anisotropy of ratio $d_x/d_y=7.5$. This means that for a soliton propagating in the $i$th direction, the PN barrier will be much less than the PN barrier for the soliton propagating in the $j$th direction, i.e., mobility has been enhanced in the direction of larger coupling. This fact will be exploited to enhance the mobility of the 2D soliton.

\begin{figure}
\centering\includegraphics[width=0.8\linewidth]{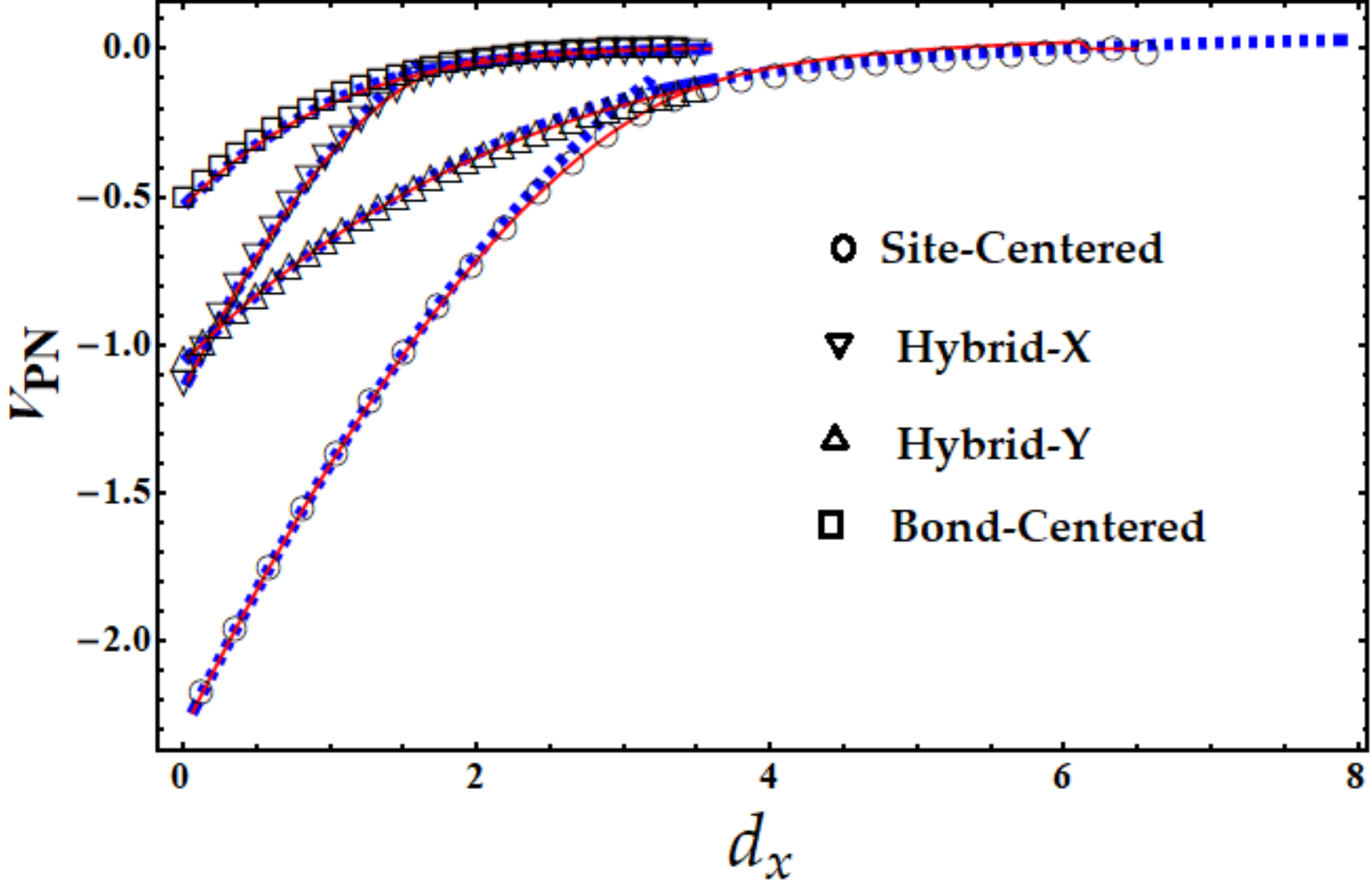}
\caption{PN barriers for the four soliton types calculated
numerically (points) and variantionaly  with a gaussian trial
function (blue) and kusp-like trial function (red). Parameters
used: $d_y=0.2$, $P=2.5$, $\gamma=4$.}\label{fig11}
\end{figure}

To further investigate the role of anisotropy on the PN potential barriers, we calculate the PN barriers for the four soliton types in terms of a wide range of anisotropy values. We perform the calculation using the numerical procedure and variational calculations described above with the two trial functions considered. The results are shown in Fig. \ref{fig11}. Both trial functions agree well with the numerical values for the whole range apart from an artifact cusp at about $d_x=3$ in the gaussian variational curve. Starting from the isotropic case ($d_x=d_y=0.2$), the PN barriers for the HX and HY solitons are seen to overlap, as expected due to the rotation symmetry of these modes. Anisotropy splits the degeneracy with the HX having smaller PN barrier indicating higher mobility than the HY soliton. Of course this would have been reversed had we taken $d_y>d_x$. As mentioned above, the SC soliton has the deepest PN barrier and the BC soliton has the shallowest barrier, thus the former being the least mobile and the later being the most mobile. Interestingly, the SC and the HY curves merge for $d_x>3.2$ and the BC and HX curves merge for $d_x>1.5$. The HY soliton is elongated in the $j$th direction. With $d_x\gg d_y$, the profile of this soliton tends to be more isotropic, hence approaching the SC profile. On the other hand, with large anisotropy the profile of the BC soliton approaches that of the HX soliton. Thus, for large anisotropy, the four soliton types reduce to two, namely those of the 1D case. In conclusion, this figure gives an idea about the anisotropy ratio at which the solitons become effectively one dimensional, at least with respect to their PN barriers.

For completeness and for the sake of comparison, we perform the variational calculation again with a the kusp-like exponential trial function
\begin{eqnarray}
\psi_{i,j}^{e}=A \,e^{-\frac{|i-n_1|}{\eta_1}-\frac{|j-n_2|}{\eta_2}+\mathit{i}v_1(i-n_1)+\mathit{i}v_2(j-n_2)}\label{expo}.
\end{eqnarray}
Due to the presence of the absolute-value function, the summations in the lagrangian can not be performed unless $n_1$ and $n_2$ are identified in advance as either integers, half integers, or any other value between two consecutive integers \cite{5,usamapn}. Having set $n_1$ and $n_2$, one can not calculate the energy functional and PN potential in terms of $n_1$ and $n_2$, as was the case with the gaussian trial function. Instead, we have to specify in advance the soliton type before calculating the energy functional. Consequently, this trial function leads to only the PN barrier heights but not the PN profile. Nonetheless, previous works have calculated the profile of the PN potential in terms of a variable, $0\le\chi\le1$, equal to the soliton location with respect to the closest site. In such a case the absolute-value function can indeed be treated analytically \cite{5}. This will give the profile of the PN potential within one cell of the periodic structure of the PN potential. For the purposes of studying the centre-of-mass dynamics of the solitons' motion across the waveguides, this will not be sufficient and the continued PN profile will be need, as already obtained by the above gaussian trial function. The kusp-like trial function will however provide an account for the solitons widths and profile that we can use to compare the results of the gaussian trial function with.
The trial functions corresponding to the four types of 2D soliton are in this case written as
\begin{equation}
\Psi_{i,j}^{\rm SC}=A\, e^{-\frac{\lvert i-L/2\rvert}{\eta_1}-\frac{\lvert j-L/2\rvert}{\eta_2}},
\end{equation}
\begin{equation}
\Psi_{i,j}^{\rm BC}=A\, e^{-\frac{\lvert i-L/2-1/2\rvert}{\eta_1}-\frac{\lvert j-L/2-1/2\rvert}{\eta_2}},
\end{equation}
\begin{equation}
\Psi_{i,j}^{\rm HX}=A\, e^{-\frac{\lvert i-L/2-1/2\rvert}{\eta_1}-\frac{\lvert j-L/2\rvert}{\eta_2}},
\end{equation}
\begin{equation}
\Psi_{i,j}^{\rm HY}=A\, e^{-\frac{\lvert i-L/2\rvert}{\eta_1}-\frac{\lvert j-L/2-1/2\rvert}{\eta_2}}.
\end{equation}
The energy functional, which is equivalent to the PN potential in this case, is then calculated by substituting these trial functions in Eq. (\ref{efunc})
\begin{align}
V_{\rm PN}^{\rm SC}&=d_x\left(1-\,\text{sech}\left(\frac{1}{\text{$\eta_1$}}\right)\right)+d_y\left(1-\text{sech}\left(\frac{1}{\text{$\eta_2$}}\right)\right) \nonumber \\ &-  \frac{1}{64}\gamma\,\text{P}\left[\left(\sinh\left(\frac{3}{\text{$\eta_1$}}\right)-\sinh\left(\frac{1}{\text{$\eta_1$}}\right)\right)
\text{sech}^3\left(\frac{1}{\text{$\eta_1$}}\right) \right. \nonumber \\ & \left. \left(\sinh \left(\frac{3}{\text{$\eta_2$}}\right)-\sinh
\left(\frac{1}{\text{$\eta_2$}}\right)\right)
\text{sech}^3\left(\frac{1}{\text{$\eta_2$}}\right)\right],
\end{align}

\begin{align}
V_{\rm PN}^{\rm BC}&=d_x\left(1-\text{sech}\left(\frac{1}{\text{$\eta_1$}}\right)
\right)+d_y\left(1-e^{-1/\text{$\eta_2$}}-e^{-1/\text{$\eta_2$}} \right. \nonumber \\ & \left. \sinh \left(\frac{1}{\text{$\eta_2$}}\right)\right)-\frac{1}{16} \gamma\,\,\text{P}\left[\,\tanh^3 \left(\frac{1}{\text{$\eta_1$}}\right) \tanh
\left(\frac{1}{\text{$\eta_2$}}\right) \right. \nonumber \\ &+ \left. \tanh\left(\frac{1}{\text{$\eta_1$}}\right) \tanh\left(\frac{1}{\text{$\eta_2$}}\right)\right],
\end{align}

\begin{align}
V_{\rm PN}^{\rm HX}&=d_x\left(1-e^{-1/\text{$\eta_1$}}-e^{-1/\text{$\eta_1$}}\sinh
\left(\frac{1}{\text{$\eta_1$}}\right)\right) \nonumber \\ 
&+ d_y\left(1-\text{sech}
\left(\frac{1}{\text{$\eta_2$}}\right)\right)\nonumber \\
&+\frac{1}{32}\,\gamma\,\, \text{P}\,\left[\tanh\left(\frac{1}{\text{$\eta_1$}}\right)
\tanh\left(\frac{1}{\text{$\eta_2$}}\right)\text{sech}^2\left(\frac{1}{\text{$\eta_2$}}\right)\right. \nonumber \\ 
& \left. -\tanh\left(\frac{1}{\text{$\eta_1$}}\right)
\sinh \left(\frac{3}{\text{$\eta_2$}}\right)
\text{sech}^3\left(\frac{1}{\text{$\eta$2}}\right)\right],
\end{align}

\begin{align}
V_{\rm PN}^{\rm HY}&=d_x\left(1- e^{-1/\text{$\eta_1$}}- e^{-1/\text{$\eta_1$}} \sinh
\left(\frac{1}{\text{$\eta_1$}}\right)\right) \nonumber  \\ &+  d_y\left(1-e^{-1/\text{$\eta_2$}}-\text{sinh}\left(\frac{1}{\text{$\eta_2$}}\right)\right)+\frac{1}{16}\gamma\,\, \text{P} \nonumber  \\ &  \tanh\left(\frac{1}{\text{$\eta_1$}}\right)\tanh\left(\frac{1}{\text{$\eta_1$}}\right).
\end{align}

Minimizing these potentials with respect to the soliton widths $\eta_1$ and $\eta_2$ gives the equilibrium widths which can then be used to plot the variational profiles, as shown by the blue lines in Figs. \ref{fig1}-\ref{fig8}. Similar to the gaussian trial function, very accurate agreement with the numerical profiles is obtained. We have also used these potentials to calculate the PN barriers, as shown in Fig. \ref{fig11} with the red lines. For most of the range of $d_x$ considered in this figure, the kusp-like and gaussian trial functions agree well with the numerical values. As mentioned above, the gaussian trial function curve shows a cusp near $d_x=3$ which we have verified as an artifact of the trial function. The kusp-like trial function does not suffer from such an artifact and continues smoothly across the numerical points at this region. On the other hand, for larger $d_x$, the gaussian trial function seems to fit the numerical points better than the kusp-like trial function.

\section{Stability phase diagram in terms of anisotropy}\label{stab_sec}
\begin{figure}
    \includegraphics[width=1.0\linewidth]{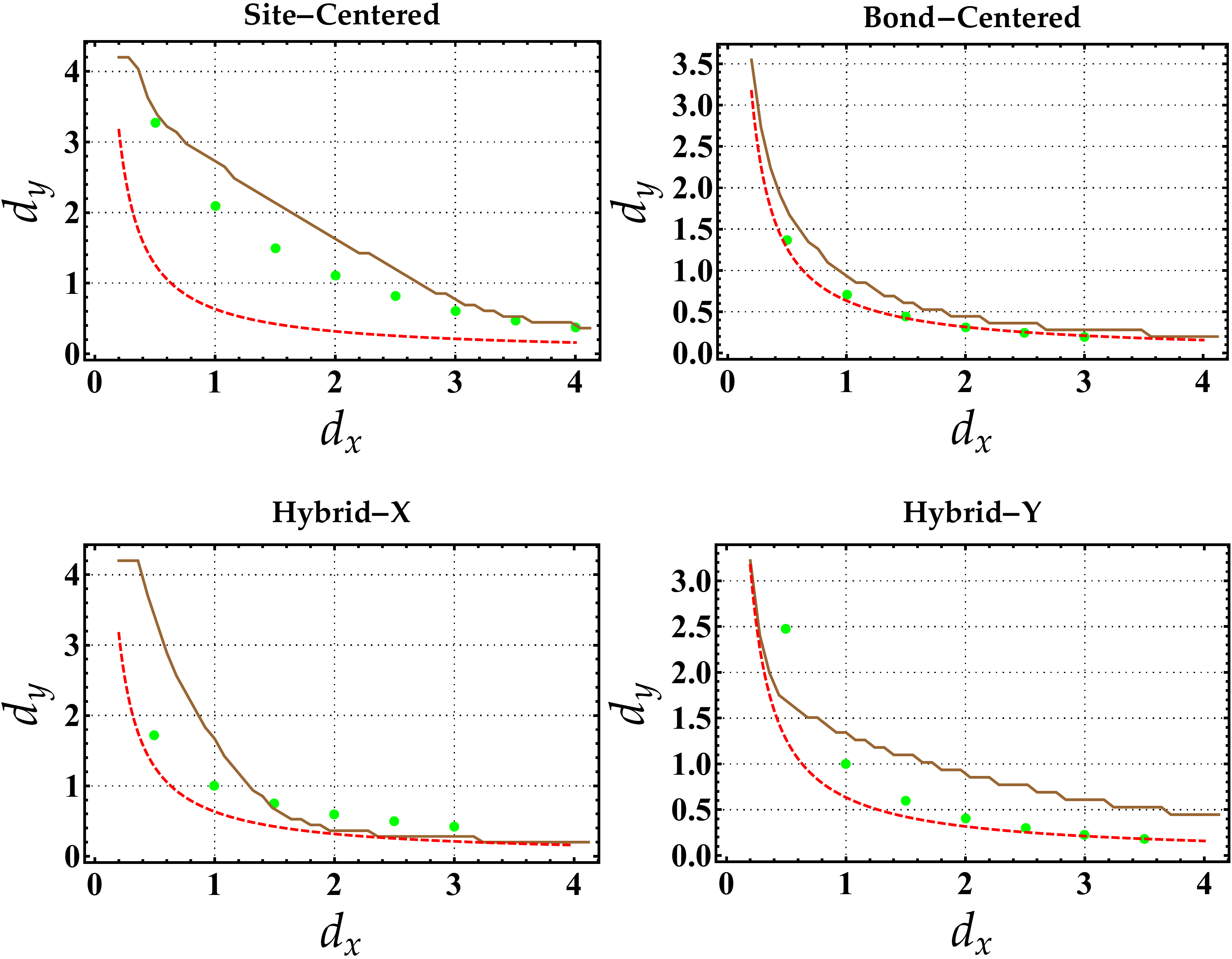}
    \caption{Stability phase diagram of 2D solitons using numerical (points) and variational calculations (solid lines). The red dashed curve corresponds to the approximate formula (\ref{eqapprox}). Parameters used $P=2.5$, $\gamma=4$.}\label{fig12}
\end{figure}

We found a limit on the anisotropy value above which stable 2D solitons do not exist; they simply decay. This was found first numerically, where we have recorded the critical values of anisotropy for which  the 2D solitons are on the border of stability. This is shown in Fig. \ref{fig12} with points for the four soliton types. The curves show a border of stability where $d_x$ and $d_y$ are inversely related to each other. The two dimensional solitons are stable for anisotropies below this border line and are unstable above it. This behaviour can be accounted for using a variational calculation. The stability region is where the PN potential $V_{\rm PN}(n_1,n_2,\eta_1,\eta_2)$, given by Eq. (\ref{PNP}), does have a minimum in terms of $\eta_1$ and $\eta_2$. Once this minimum is lost, the width of the soliton, according to the variational calculation, diverges. In mathematical terms, the condition is written as
\begin{equation}
\frac{\partial V^{\rm SC}_{\rm PN}(\eta_1,\eta_2)}{\partial \eta_1}\ne0,\, {\rm or}\, \frac{\partial V_{\rm PN}^{\rm SC}(\eta_1,\eta_2)}{\partial \eta_2}\ne0,\,{\text {for all}}\,\,\eta_{1,2},
\label{eqcond}
\end{equation}
and similarly for the other soliton types.
This is equivalent to the observed decay in the numerical solution. To verify this, we have calculated the border line at which the minimum in the PN potential starts to disappear, which is plotted with solid curves in Fig. \ref{fig12}. Quantitatively, there is a good agreement with the numerical border.

A crude but simple analytic formula for the stability border can be obtained from the variational calculation in the large $\eta_1$ and $\eta_2$ limit which is equivalent to the condition of decaying soliton, as mentioned above. In this limit, the zeroth order expansion of conditions defining the border, read
\begin{equation}
\frac{-4d_x+P\gamma\,\eta_1/(\pi\,\eta_2)}{4\eta_1^3}=0,\hspace{1cm}\frac{-4d_y+P\gamma\,\eta_2/(\pi\,\eta_1)}{4\eta_2^3}=0,
\end{equation}
which are solved for
\begin{equation}
d_x=\frac{\gamma^2\,P^2}{16\pi^2\,d_y}
\label{eqapprox}
\end{equation}
confirming the inverse relation between $d_x$ and $d_y$ on the stability border. This is a very rough formula as can be seen when plotted versus the numerical and variational results in Fig. \ref{fig12} shown with the red dashed curves. There  is though a good quantitative agreement for the BC soliton. The approximation can be enhanced by taking higher order terms in the expansion.

\section{Enhanced Mobility}\label{mob_sec}

\subsection{Stability of movable 2D solitons}
\label{stability_sub1}

In this section, we investigate the role of anisotropy of the coupling coefficients on the mobility of 2D solitons. We have seen in the previous section that the PN barrier reduces in one direction for anisotropic coupling coefficients. We show here that 2D solitons will be indeed mobile in the direction of reduced PN barrier. However, there will be a critical anisotropy value above which mobility is triggered. The critical anisotropy value depends on the kick-in speed given initially to the soliton in order to move it. Therefore, investigating the mobility requires scanning the parameter space of both anisotropy ratio and kick-in speed. Since the Site-Centered  soliton is the most pinned among the four soliton types, as can be seen in Tables 1 and 2 where the PN barrier depth is the largest in magnitude, we investigate the role of anisotropy on this type of 2D solitons. The rationale is that if the most pinned soliton is turned to mobility by a certain value of anisotropy, all other types will be also mobile by the same amount of anisotropy.

\begin{figure}
    \includegraphics[width=1.0\linewidth]{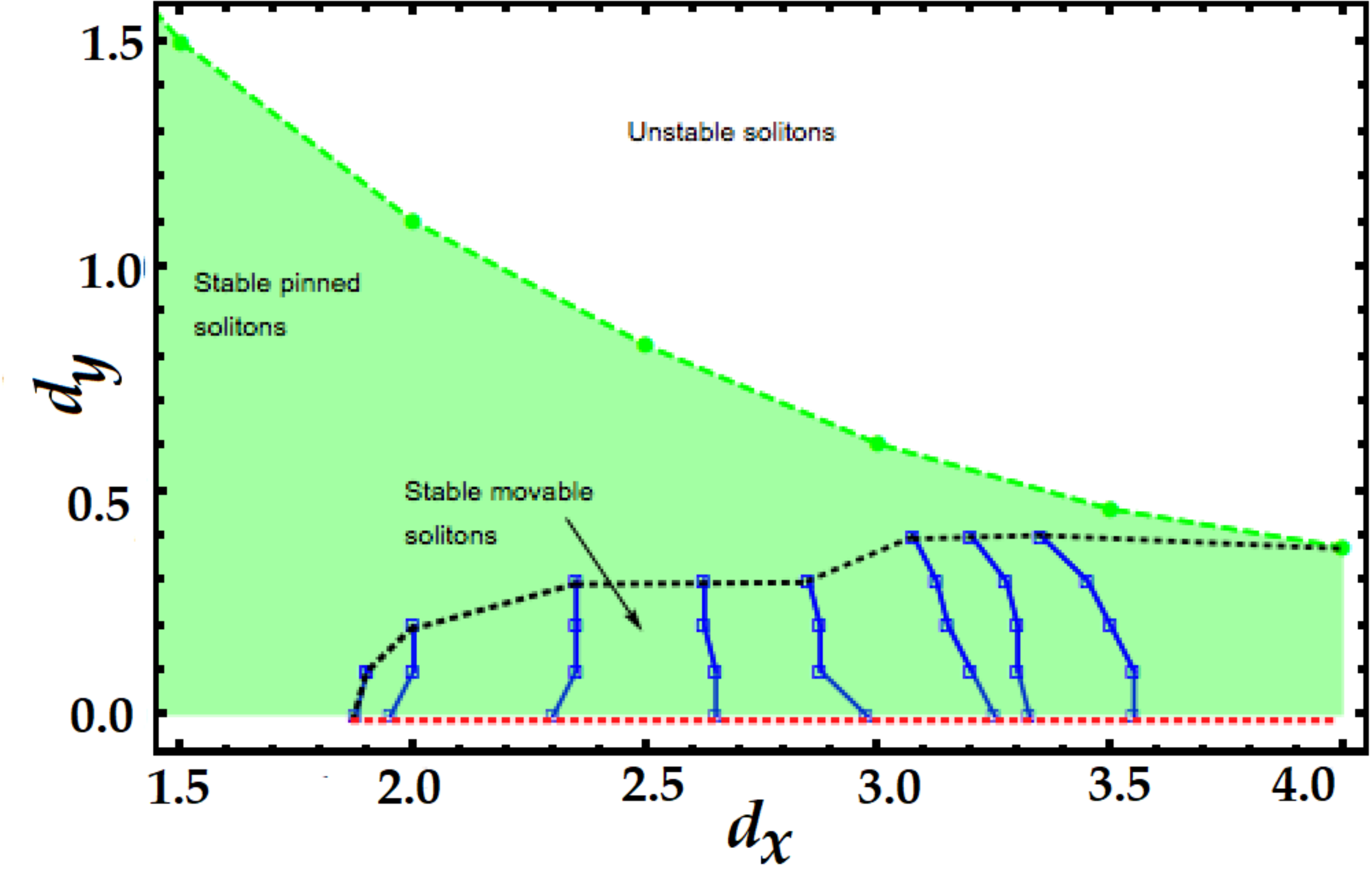}
\caption{  Phase diagram showing stability and mobility of the
Site-Centered 2D soliton (inset of Fig.\ref{fig12}) in terms of
the coupling coefficients $d_x$ and $d_y$, which define the
anisotropy ratio. Stable stationary (pinned) 2D solitons exist
only in the shaded area. Movable stable solitons exist in the area
below the dashed black line. Points (open squares) connected by
lines correspond to anisotropy thresholds for solitons mobility;
solitons are mobile only in the area to the right of each of these
lines while each line corresponds to a different initial kick-in
speed which read, starting from the right: $v_x=0.1, 0.2, 0.3,
0.5, 0.7, 1.0, 1.5, 1.85$. The red dashed line corresponds to the
1D case. Parameters used are: $\gamma=4,\, P=2.5,\, L=20$. }
 \label{fig10phase}
\end{figure}

The previously-found phase diagram of Site-Centered soliton stability, Fig.\ref{fig12}, will be the basis for our study in this section. We have performed a systematic investigation by scanning the whole parameter space of anisotropy - in the stable-solitons part of this diagram - and kick-in speed values by solving numerically Eq. (\ref{eq1}) and recording the critical values of these parameters at which the soliton starts to leave its site. We point out here that we followed this criterion for mobility with no regard to where the soliton will stop later on due to friction with the PN potential. The result of this investigation is shown in Fig.\ref{fig10phase}. We found that solitons are mobile only in the region below the  black dashed line. Outside this region, the solitons are not mobile for any nonzero value of the kick-in speed; they start decaying once they start the motion. In the mobility region, solitons also show some decay in their amplitude as they start the motion, but then keep a finite value of amplitude for a very long time (we show below how long is that time). The mobility region is also divided by the values of the kick-in speed into subregions, as shown by the blue lines with square points. For each of these lines a different kick-in speed is used with the lowest kick-in speed for the the line on the right, see the caption of the figure for details.  For a given kick-in speed, the soliton is mobile only in the region to the right of the blue line. As the anisotropy decreases, larger speed is required to move the soliton which is consistent with the above-mentioned fact of PN barrier increasing with decreasing anisotropy. There is also an upper limit on the speed above which the solitons are not mobile for any anisotropy even for the extreme case of 1D solitons ($d_y=0$). This is given by the left end of the red dashed line on the horizontal axis. The square points on this dashed line give the critical mobility kick-in speed of the 1D soliton in terms of $d_x$.

\begin{figure}
    \includegraphics[width=1.0\linewidth]{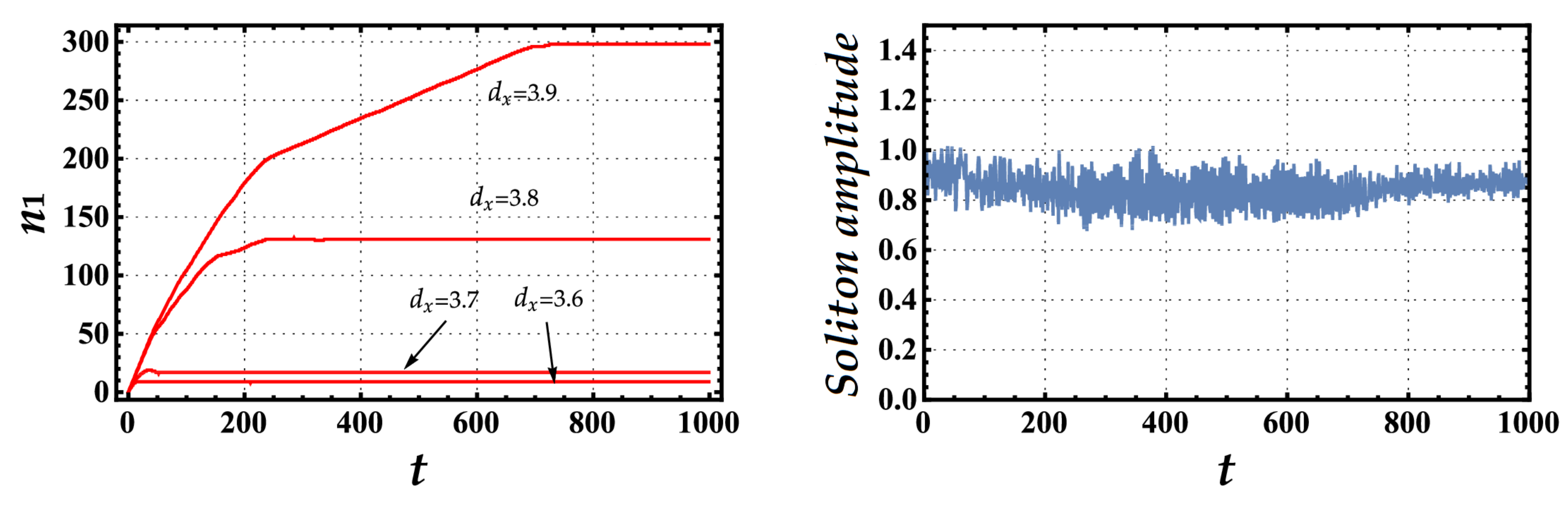}
    \caption{
     Soliton peak position (left) for four values of $d_x$, $d_y=0.1$, and initial speed $v_x=0.2$. Amplitude of the soliton is shown (right) for the case with $d_x=3.9$. Other parameters used are $\gamma=4$, $P=2.5$ and $L=30$. Initial soliton position is at $(n_1,n_2)=(15,10)$.
    }\label{fig0mobility}
\end{figure}

To show a specific case of mobility, we plot in Fig.\ref{fig0mobility} the trajectory of the soliton and its amplitude for four values of $d_x$ with a fixed kick-in speed, namely $d_x=3.6,\,3.7,\,3.8,\,3.9$, $d_y=0.1$, and $v_x=0.2$. Since the soliton is kicked only in the $x$-direction ($v_y=0$), it does not move in the $y$-direction therefore we show only the evolution of $x$-component of the soliton position, namely $n_1$ while $n_2$ remains a constant that is equal to its initial value. We have employed periodic boundary conditions to be able to track the soliton trajectory for long times. It is clear that increasing the anisotropy enhances considerably on the soliton mobility. All curves show the common feature of dissipative motion due to radiation losses caused by the PN potential. As a result, the solitons' speed reduces to a value that such that the soliton will be ultimately pinned. However, for the large the anisotropy ratios, such as $d_x/d_y=38$ and $39$ for the upper two curves, the soliton settles at values as large as $n_1=170$ and $300$, respectively. For $d_x/d_y=36$ and $37$, the soliton gets pinned much earlier. For $d_x/d_y<36$, the soliton is completely pinned.  This gives an idea of how much anisotropy is needed to unpin the soliton. In the mobility cases, such as $d_x/d_y=39$, the amplitude of the soliton oscillates around a finite value though is slightly less than the initial value due to losses by the PN potential. This figure shows that 2D highly mobile solitons exist but with anisotropy ratio larger than one. It should be stressed here that this study was performed for the Site-Centered soliton with $d_x\gg d_y$. For the other types of soliton and anisotropy ratios, lower values of anisotropy will be required to render the soliton to mobility.

\subsection{Accelerating and routing 2D solitons}
\label{manage_sec}

It is established for discrete solitons propagating in one-dimensional waveguide arrays that an effective potential can be created by varying the strengths of the coupling coefficients across the waveguides \cite{andrey, usama}.  The profile of the effective potential is directly proportional to that of the coupling coefficients. Furthermore, the strengths of the coupling coefficients can be varied by varying the separations between the waveguides; the strength of the coupling coefficient decays exponentially with the separation between waveguides \cite{exptcouplings}.

\begin{figure}
         \includegraphics[width=1.05\linewidth]{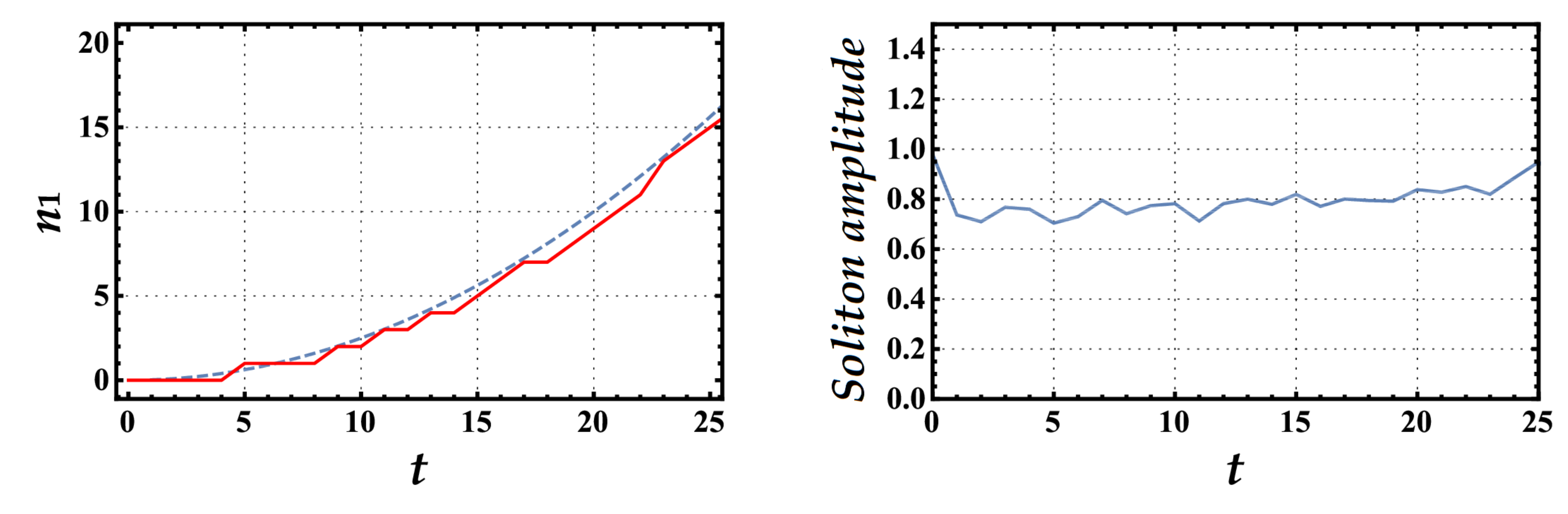}\\
         \includegraphics[width=1.05\linewidth]{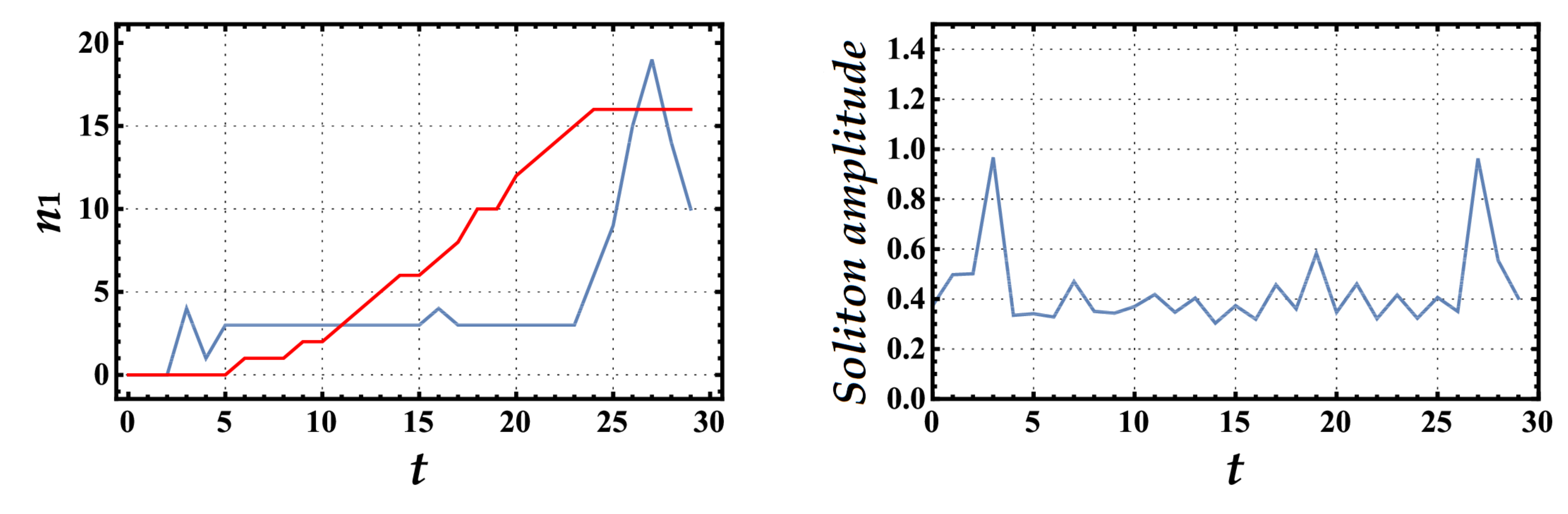}
   \caption{Upper panel: Accelerating a soliton using linearly increasing coupling $d_x=3.9+0.05\,i$ (left). The dashed curve corresponds to the `force', $F^{eff}=0.025\,i^2$. Soliton amplitude is shown to oscillate around a constant finite value (right). Parameters used: $d_y=0.1$, $\gamma=4$, $P=2.5$, $L=30$, initial soliton position $(i,j)=(15,10)$.
Lower panel: Soliton trajectory and amplitude along the 3-branches track defined by Eqs. (\ref{branch1}-\ref{branch3}). On the left subfigure, red corresponds to $n_1$ and blue corresponds to $n_2$. Parameters used: $\gamma=4$, $P=2.5$, $L=30$, initial soliton position $(i,j)=(15,10)$. }\label{fig00mobility}
\end{figure}

Combining this fact with that found in the previous section, namely 2D solitons being mobile in anisotropic waveguide arrays, one can design 2D waveguide profiles to control the flow of the solitons in two dimensions. In the following, we demonstrate this idea by two examples: accelerating a soliton along one direction and routing the soliton from one direction to the other. In the first example, we show that a linearly increasing strength of the coupling coefficients in one direction amounts to a linear effective potential which results in an accelerated soliton in that direction. Specifically,
\begin{eqnarray}
d_x(i,j)&=&3.9+0.05i,\nonumber\\
d_y(i,j)&=&0.1.
\end{eqnarray}
It is noted that our specific choices of the values of $d_y=0.1$ and initial value of $d_x=3.9$ is based on our previous finding that the 2D soliton will be highly mobile with these parameters, see Fig.\ref{fig0mobility}. The time evolution of the soliton is performed in two steps. First, we prepare the initial soliton by solving Eq. (\ref{eq1}) with $d_x=3.9$ and $d_y=0.1$, as described in section \ref{num_sec}. This will result in a stable and stationary soliton profile that is elongated along the $i$-direction and hence has high flexibility to move in that direction. The second step is to evolve this stationary soliton by inserting the above index-dependent coefficients in Eq.(\ref{eq1}). Effectively, this will be equivalent to evolving a stationary soliton in a 2D waveguide array with constant coefficients and an effective potential in the $i$th direction, namely $V^{eff}(i)\propto -d_x(i)$. The resulting numerical simulation supports this description, as shown in Fig.\ref{fig00mobility} where the peak position is indeed being accelerated along the $i$th direction. The width and peak height of the soliton also remain constant on the average which shows that the soliton preserved its integrity in such an inhomogeneous medium. Furthermore, one can see that the soliton is being affected by a constant `force' as a result of the effective potential, namely $F^{eff}\propto-d V^{eff}/di=0.05i$. The trajectory of objects moving by a constant force is parabolic. The peak position should thus follow the trajectory $n_1(i)=\int F^{eff}di=(n_1)_0+0.025i^2$, where $(n_1)_0$ is the initial position. This prediction fits perfectly with the numerical result for the trajectory, as Fig.\ref{fig00mobility} shows.

\begin{figure}
    \includegraphics[width=1.0\linewidth]{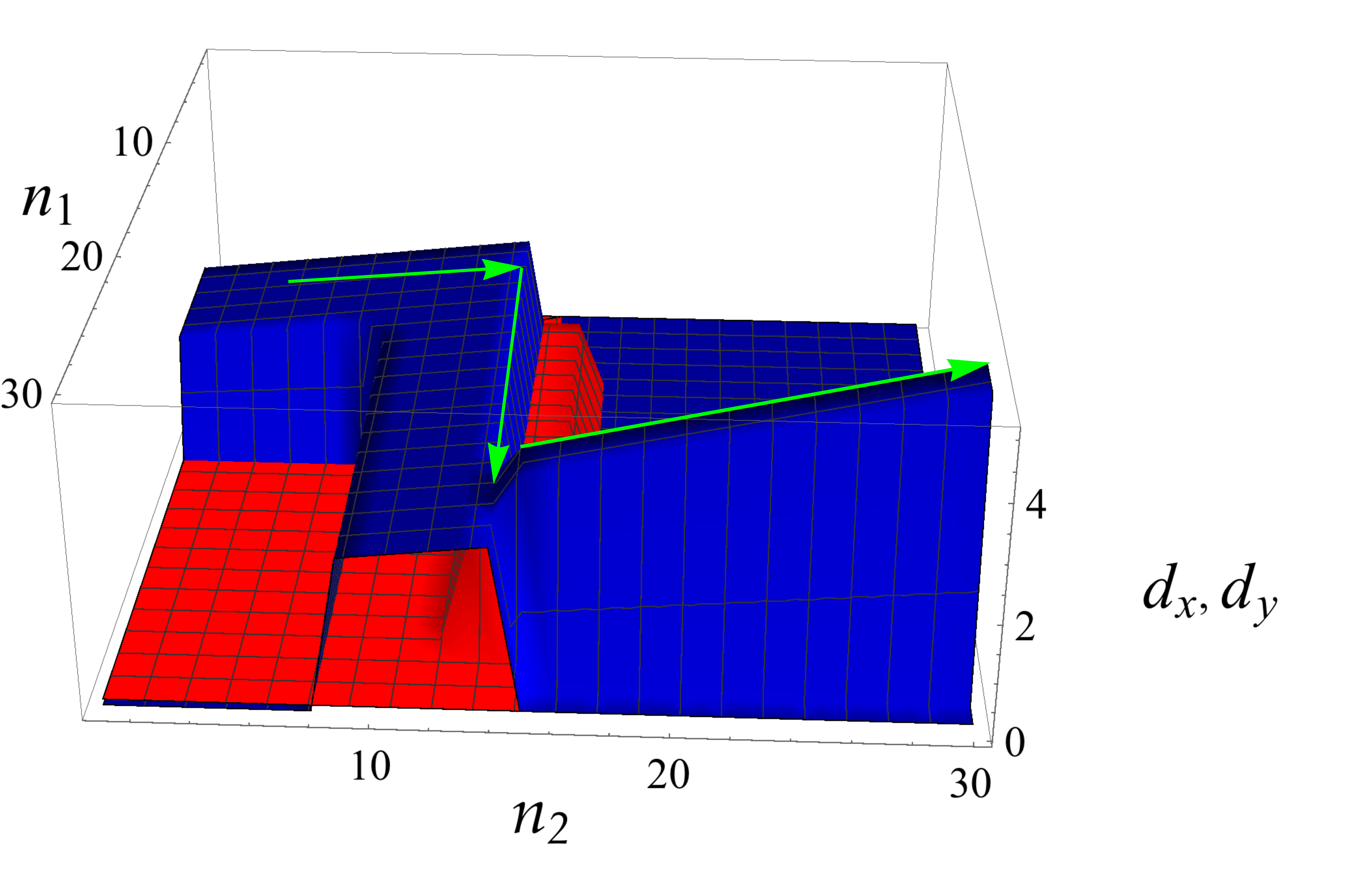}
    \caption{
    Strengths of the coupling coefficients along the three branches defined by Eqs. (\ref{branch1}-\ref{branch3}). Blue surface refers to $d_y(i,j)$ and red surface refers to $d_x(i,j)$. The arrows show the direction of the soliton trajectory.
    }
    \label{fig3mobility}
\end{figure}

We exploit this possibility of accelerating solitons in our second example where we also route the soliton performing two $90^{\rm o}$-bends in its trajectory.  We start with a 2D waveguide array with isotropic homogeneous couplings, $d_x(i,j)=d_y(i,j)=0.1$ everywhere except along certain paths where we modulate the couplings such that the soliton will be accelerated along these paths. The path we design is made of three branches: Branch 1: the soliton is accelerated along the $j$th direction using this profile
\begin{equation}
d_y(i,j)=2.2+0.05j,\hspace{1cm} j\le8,\,\, 8\le i \le 14,
\label{branch1}
\end{equation}
then in Branch 2 which starts at the end of Branch 1, the soliton is accelerated in the $i$th direction with this profile
\begin{eqnarray}
d_x(i,j)&=&1.1+0.03i,\hspace{1cm} 13\ge j \le16,\,\, 8\le i \le 26,\nonumber\\
d_y(i,j)&=&2.2+0.05j,\hspace{1cm} 8<j<15,\,\, 8\le i \le 26.\nonumber\\
\label{branch2}
\end{eqnarray}
Finally, in Branch 3, the soliton is again accelerated in the $j$th direction with the profile
\begin{equation}
d_y(i,j)=2.0+0.1j,\hspace{1cm} j\ge15,\,\, 26\le i \le 28.
\label{branch3}
\end{equation}
Outside these three branches the values of $d_x(i,j)$ and $d_y(i,j)$ take their default value of 0.1. The strengths of the coupling coefficients along these three branches are shown in Fig.\ref{fig3mobility}.

\begin{figure}
     \includegraphics[width=1.0\linewidth]{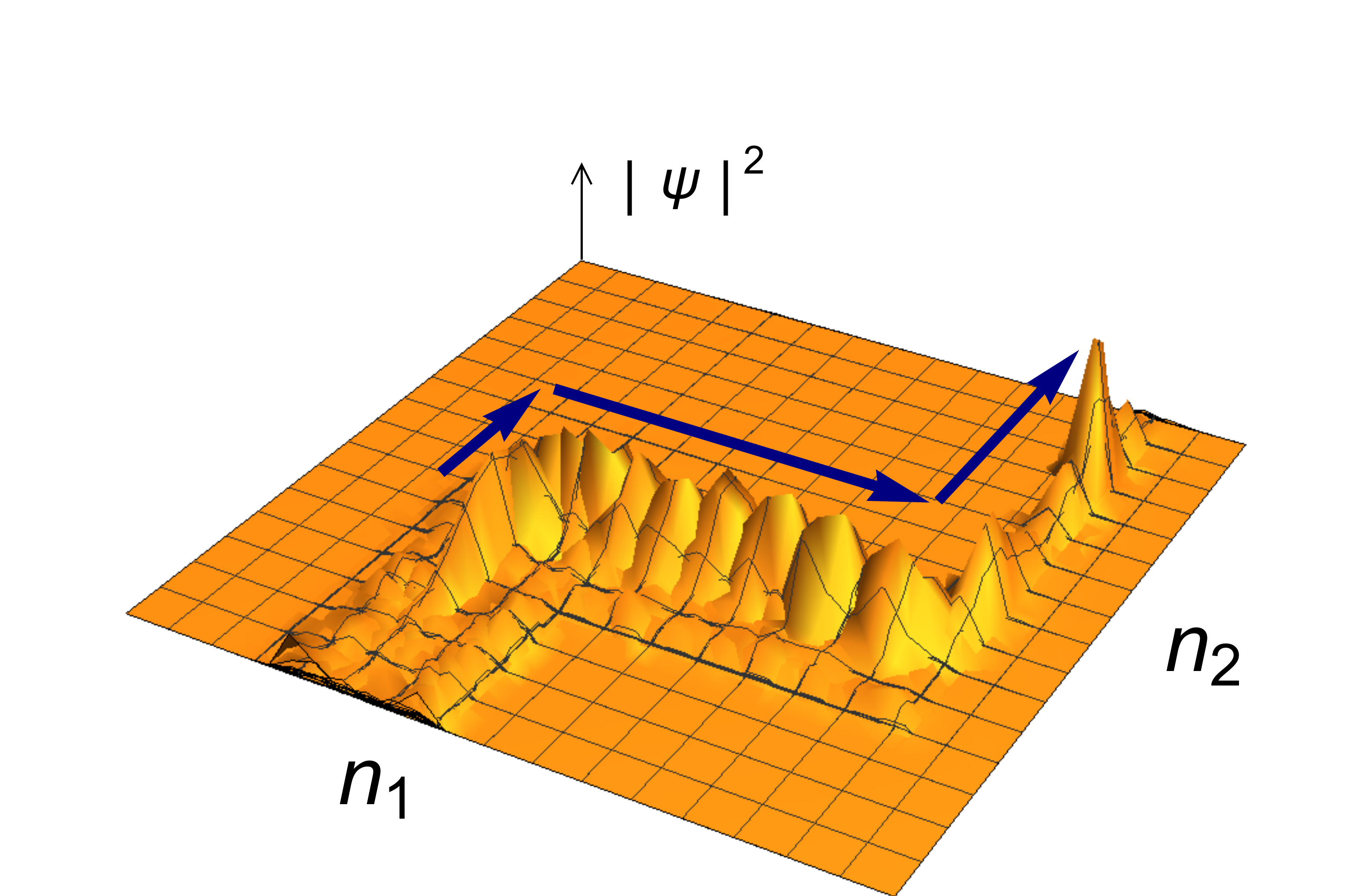}
    \caption{Time evolution of a 2D soliton with couplings' profile given by Eqs. (\ref{branch1}-\ref{branch3}). Initially, the soliton is located at $(i,j)=(10,15)$.
    The arrows show the direction of the motion. }\label{fig2mobility}
\end{figure}

Similar to the previous example, we first prepare an elongated stationary soliton with high anisotropy, $d_y=8$ and $d_x=0.1$. Then we use  Eq. \ref{eq1} to evolve this soliton using the above coupling profiles. The resulting dynamics shows the 2D soliton indeed following the designed path, as shown in Fig.\ref{fig2mobility}. The trajectory and width of the soliton are plotted in Fig.\ref{fig00mobility}. It is clear from these two figures that while there is a reduction from the initial amplitude of the soliton due to radiation losses, the soliton keeps its integrity by preserving an average finite width and amplitude along the three segments of the path.  It can also be noted that the soliton profile is compressed at the two turning points connecting the different branches. At these points the soliton is forced to react to the sudden change in the anisotropy of the waveguides by modulating its widths and amplitude.

\begin{figure}
    \includegraphics[width=1.0\linewidth]{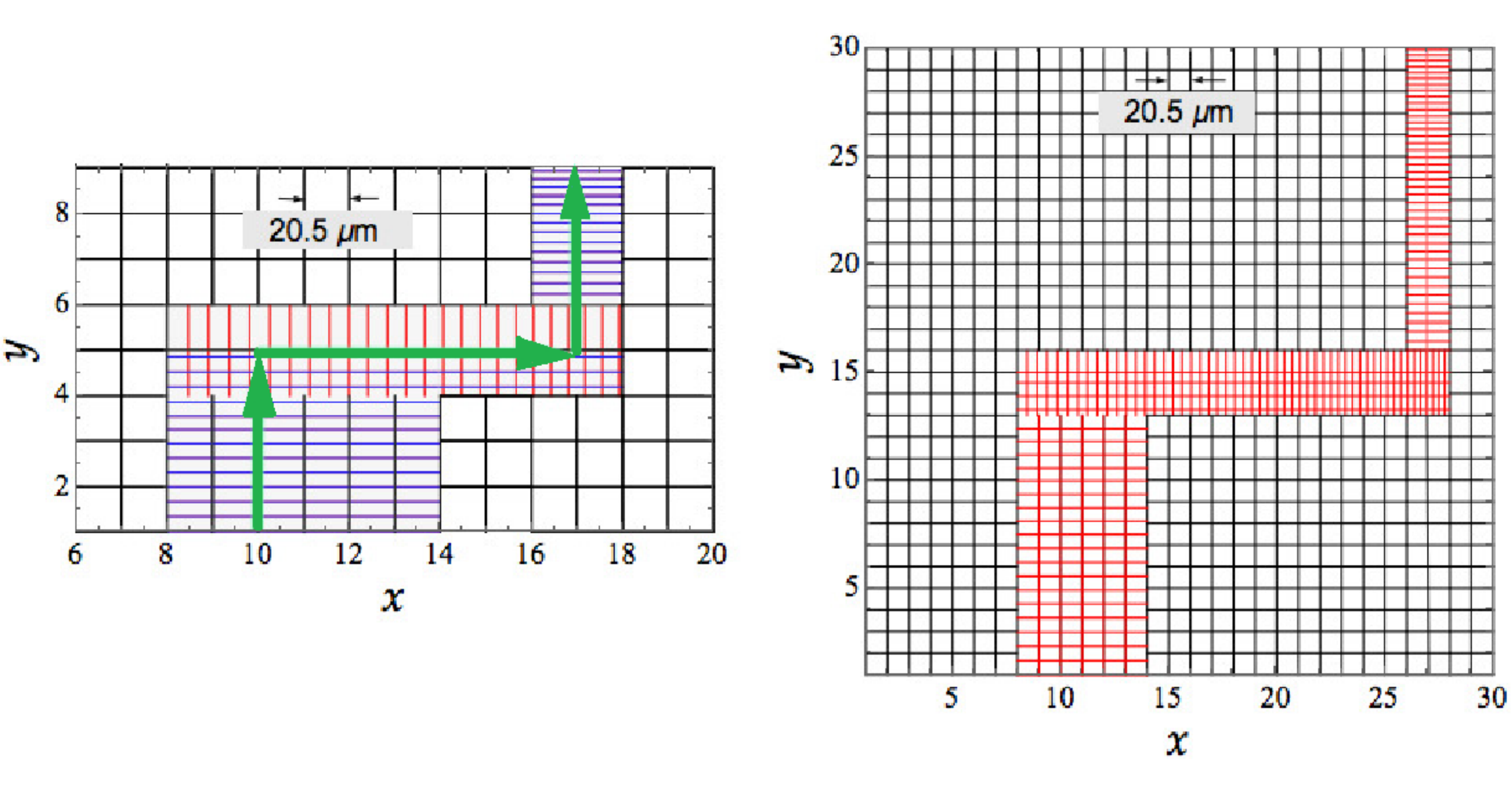}
    \caption{ Waveguides separations profile corresponding to the three-branches trajectory defined by Eqs. (\ref{branch1}-\ref{branch3}).
    Waveguide sites are at the intersection between horizontal and vertical lines. Left: gaps between the three branches are filled by stacking them to each other. The arrows indicate the direction of soliton's motion. Right: gaps between the branches are filled by adding more waveguides.
    }
       \label{fig5mobility}
\end{figure}

The profile of coupling coefficients, Eqs. (\ref{branch1}-\ref{branch3}), leading to the three branches trajectory can be realized by modulating the waveguides' separations. It is found experimentally that the strength of the coupling coefficients decay exponentially with their separation, see Fig.3b of Ref.\cite{exptcouplings}. Fitting the experimental data, for the wavelength $543$nm, with an exponential law, we find the following relation
\begin{equation}
d=d_0\,e^{-\frac{r-r_0}{a_0}}
\end{equation}
with $r_0=14{\rm\mu}$m, $a_0=4.3{\rm\mu}$m, $d_0=0.45$cm$^{-1}$. Here, $r$ is the separation between two consecutive waveguides and $d$ is the coupling strength between them. Inverting this relation to express the separation in terms of coupling, we obtain
\begin{equation}
r=14-4.3\,\log{\left(\frac{d}{0.45}\right)}.
\end{equation}
There is a maximum coupling strength, $d=11.67$cm$^{-1}$, corresponding to zero separation.  This value is well above the coupling strengths used in the present paper and thus our coupling strengths can be realized with finite separations between the waveguides. Using this relation, the above coupling strengths profile for the three-branches trajectory, Eqs. (\ref{branch1}-\ref{branch3}), can be `translated' into a waveguide separations profile, as shown in Fig.\ref{fig5mobility}. Since we are accelerating the solitons along the three branches which amounts to a linearly increasing coupling strength, the separations between the waveguides will be decreasing. As a result the total length of the branch will be less than its equivalent length for a uniform coupling profile. This will create a gap between the branches. We propose two methods of filling these gaps. One method is to shift the waveguides as a whole and stack them to each other, as shown by the left subfigure in Fig.\ref{fig5mobility}. Alternatively, we may add more waveguides along a branch by extrapolating the separations law along that branch, as shown by the right subfigure.

\section{Conclusions}
\label{conc_sec}
There are four main conclusions of the present work. Firstly, we have shown that it is possible to enhance the mobility of 2D discrete solitons by breaking the isotropy in the coupling coefficients. Mobility will be enhanced in the direction with larger coupling coefficients. Second, we have obtained the 2D PN potential using a variational calculation with a gaussian trial function. Third, we have obtained a phase diagram showing regions of stability and subregions of mobility in terms of the anisotropy of the waveguide array. Fourth, we have shown that it is possible to guide and route 2D solitons by designing tracks with anisotropic coupling coefficients and modulated separations.

We started by showing that stationary solitons exist in anisotropic waveguide arrays as they do in isotropic waveguides but with an important difference. The difference is that in the isotropic waveguides only three fundamental stationary soliton types exist, as shown in Figs. \ref{fig1}-\ref{fig4}. In the present anisotropic case, four types exist: Site-Centered and Bond-Centered,  Hybrid-X, and  Hybrid-Y solitons, as shown in Figs. \ref{fig5}-\ref{fig8}. So, what used to be one type in the isotropic case, namely the hybrid soliton, degenerates in the anisotropic case into the Hybrid-X, and  Hybrid-Y solitons. These results were obtained from the numerical solution of the governing equation, Eq. (\ref{eq1}), of 2D solitons. We have also laid down the framework of a two-dimensional variational calculation that predicts the existence of the four stationary soliton types and accounts accurately to their equilibrium widths, as shown by Figs. \ref{fig1}-\ref{fig8}.  We have used a gaussian variational trial function, Eq. (\ref{gaussian}), with six variational parameters corresponding to the two components of the soliton position, width, and velocity. We have also used kusp-like variational function, Eq. (\ref{expo}), for the purpose of comparison. The advantage of the gaussian trial function is that the extended PN potential can be obtained, as shown in Figs. \ref{fig9}-\ref{fig10}. In addition, the variational calculation reproduces accurate values of the PN barrier hight for the four soliton types in comparison with the numerical values, as shown in Fig. \ref{fig11}.

The stability of 2D solitons against collapse was then investigated versus the anisotropy ratio using both numerical and variational calculations. Good agreement was obtained between both calculations for the phase diagram showing stability region versus the coupling coefficients $d_x$ and $d_y$. This was performed for the four soliton types, as shown in Fig. \ref{fig12}. Then we investigated the stability of the most pinned soliton, namely the Site-Centered soliton, in terms of the initial kick-in speed. This resulted in the mobility phase digram shown by Fig. \ref{fig10phase}.

Controlling the trajectory of the 2D solitons was then demonstrated by two examples. In the first example, we have shown that it is possible to accelerate the soliton along a track where the coupling strength is increasing, as shown in Fig. \ref{fig0mobility}. In the second example, we have designed a track composed of three segments along which the soliton is being accelerated by the same method a in the first example, see Fig. \ref{fig3mobility}. This resulted in the soliton following the designed path that included two $90^{\rm o}$-bends, as shown in Figs. \ref{fig00mobility} and \ref{fig2mobility}. Finally we have calculated the separations between the waveguides needed to perform the predicted guided trajectory of 2D solitons. This was based on the experimental calibration of the coupling strength decay in terms of waveguides separations \cite{exptcouplings}, as shown in Fig. \ref{fig5mobility}.

We have focused in the present work on the mobility of the most pinned 2D soliton, namely the Site-Centered soliton. However, one may also consider other types of soliton where even better mobility is expected to be obtained. In addition, it will be interesting to study the mobility of the anisotropic solitons, such as the HX soliton, along the different directions including the horizontal, vertical and diagonal directions. High contrast is expected to be observed in this case. One may also consider performing all-optical operations using the routing mechanism described here. One of the most looked for goals in this respect is to achieve the function of a transistor which requires 2D waveguides.

\end{document}